\documentclass[a4paper,11pt]{article}

\usepackage{jinstpub} 
\usepackage{placeins}
\usepackage{lineno}
\usepackage{graphicx}
\usepackage{todonotes}
\usepackage{caption} 
\usepackage{subcaption} 
\usepackage{mathdots}

\title{\boldmath NEXT-CRAB-0: A High Pressure Gaseous Xenon Time Projection Chamber with a Direct VUV Camera Based Readout}
\collaboration{The NEXT Collaboration}

\author[1,a]{N.K.~Byrnes,\note[a]{Corresponding Author, nicholas.byrnes@uta.edu}}
\author[1,a]{I.~Parmaksiz,\note[a]{Corresponding Author, ilker.parmaksiz@mavs.uta.edu}}
\author[2]{C.~Adams,}
\author[1]{J.~Asaadi,}
\author[1]{J~Baeza-Rubio,}
\author[2]{K.~Bailey,}
\author[3]{E.~Church,}
\author[4]{D.~González-Díaz,}
\author[2]{A.~Higley,}
\author[1]{B.J.P.~Jones,}
% ...
\affiliation[1]{
Department of Physics, University of Texas at Arlington, Arlington, TX 76019, USA}
\affiliation[2]{
Argonne National Laboratory, Argonne, IL 60439, USA}
\affiliation[3]{
Pacific Northwest National Laboratory (PNNL), Richland, WA 99352, USA}
\affiliation[4]{
Instituto Gallego de F\'isica de Altas Energ\'ias, Univ.\ de Santiago de Compostela, Campus sur, R\'ua Xos\'e Mar\'ia Su\'arez N\'u\~nez, s/n, Santiago de Compostela, E-15782, Spain}
% ...

\emailAdd{nicholas.byrnes@uta.edu}
\emailAdd{ilker.parmaksiz@mavs.uta.edu}

\abstract{
The search for neutrinoless double beta decay ($0\nu\beta\beta$) remains one of the most compelling experimental avenues for the discovery in the neutrino sector.
Electroluminescent gas-phase time projection chambers are well suited to $0\nu\beta\beta$ searches due to their intrinsically precise energy resolution and topological event identification capabilities. 
Scalability to ton- and multi-ton masses requires readout of large-area electroluminescent regions with fine spatial resolution, low radiogenic backgrounds, and a scalable data acquisition system.
This paper presents a detector prototype that records event topology in an electroluminescent xenon gas TPC via VUV image-intensified cameras. 
This enables an extendable readout of large tracking planes with commercial devices that reside almost entirely outside of the active medium. 
Following further development in intermediate scale demonstrators, this technique may represent a novel and enlargeable method for topological event imaging in $0\nu\beta\beta$. 
}

\begin{document}
\maketitle{}
\flushbottom

\section{Introduction}
\label{sec:intro}

The nature of the neutrino has been a subject of theoretical and experimental inquiry since its inception in the early days of particle physics \cite{earlynu}. 
The confirmation of massive neutrino states \cite{bilenky1987massive,Bandyopadhyay:2002xj,SNO:2002tuh,Super-Kamiokande:1998kpq} invites extensions to the standard model that violate lepton number, in such a way that neutrinos are their own anti-particles \cite{Dib:2000ce,Faessler:1998sba}. 
The most experimentally accessible avenue to search for this process - known to be exceedingly rare, if allowed - is through the observation of neutrinoless double beta decay ($0\nu\beta\beta$) \cite{jones2021physics, dolinski2019neutrinoless, giunti2022report}. 

In a nucleus undergoing an ordinary double beta decay event, two nucleons simultaneously beta-decay with the emission of four leptons: two electrons and two electron anti-neutrinos~\cite{PhysRev.48.512}. 
Experimental observation of this process has been made in a handful of nuclei where single beta decay is energetically forbidden and the double beta decaying isotope has high natural abundance.
Measured half-lives for $2\nu\beta\beta$ are in excess of $10^{18}$ years \cite{universe6100159}, with the half life of $^{136}$Xe - the nucleus of interest for this work - measured at approximately $2.17 \times 10^{21}$ years in liquid xenon \cite{Ackerman_2011,KamLAND-Zen:2012vpv} and confirmed via topological event identification and direct background subtraction in xenon gas \cite{Novella_2022}. 
If and only if the neutrino is its own antiparticle, a second kind of double beta decay with no neutrinos is possible. 
Experimental observations have confirmed the theoretical predictions that the double decay in the zero-neutrino mode is much rarer than the two-neutrino mode, with a half-life lower limit established to be in the range of $10^{26}$ years in xenon \cite{KamLAND-Zen:2022tow,GERDA:2020xhi}.

Neutrinoless double beta decay searches employ several strategies to achieve sensitivity. 
First, the long half-life of the decay is counterbalanced by building detectors containing as much of the target isotope mass as is feasible, within the constraints of engineering, underground space, funding, and isotopic procurement. 
Second, experiments are operated deep underground to minimize incidental backgrounds originating from cosmic-ray activity. 
Third, radiogenic backgrounds near the decay Q-value ($Q_{\beta \beta}$) are minimized as much as possible through the combination of careful screening of detector components as well as application of selection criteria to discriminate signal from background. 
In the NEXT~\cite{ Alvarez:2012yxw, Alvarez:2012haa, Novella:2019cne, Martin-Albo:2015rhw,adams2021sensitivity} program of time projection chambers (TPCs), extended beta tracks in high pressure gaseous xenon (HPGXe) are imaged via their secondary scintillation light generated through the electroluminescence (EL) process. 

When a charged particle passes through a gaseous TPC detector it ionizes and excites the surrounding medium. The de-excitation of xenon results in isotropic emission of photons. 
The central wavelength for these scintillation photons has been reported as 170 $\pm 5.9$~nm \cite{170nm},and 172 $\pm 5.5 $~nm \cite{172nm} at 10 bar xenon. 
The initial recombination and de-excitation produce photons (S1) which encode the interaction time of an event. 
The active region of the detector is subjected to an electric field $\mathcal{O}(100)$~V/cm, which serves to drift the ionization electrons to an EL region. 
An electric field of $\mathcal{O}(20)$~kV/cm in the EL region accelerates the electrons to sufficiently high energy between collisions that they collisionally excite the xenon atoms. This secondary scintillation light (S2) is used to establish both event energy and topology, with typical gains of order a few hundred to a thousand photons per incident electron. 

If the EL electric field is kept below the threshold for secondary ionization, avalanche fluctuations in the light yield are avoided. 
Xenon gas has a Fano factor of around 0.16 \cite{ANDERSON1979125} and the EL process provides a near fluctuation-less gain, the combination of which leads to an energy resolution that is world leading among xenon detectors~\cite{Monrabal:2018xlr}. Currently, NEXT detectors collect the photons with photomultiplier tubes (PMTs) to measure event energy. The discrimination power afforded by the event energy is presently the only known way to distinguish between $0\nu\beta\beta$ and the irreducible $2\nu\beta\beta$ background. 

The use of a gaseous medium allows also for topological imaging of a distinguishable track with a Bragg peak at its end for each beta electron. This provides a powerful handle for active background rejection, so long as the track can be captured with sufficient spatial resolution. The strength of this track recognition has been demonstrated \cite{Ferrario:2019kwg} by collecting the photons with a pixelized plane of silicon photomultipliers (SiPMs) to provide X and Y coordinates, with charge arrival time providing the Z component to yield a 3-dimensional track. 
The blurring effects of diffusion and smearing by the wavelength shifter in the EL region have been shown to be significantly reduced by the addition of helium gas to the detector~\cite{Felkai:2017oeq, McDonald:2019fhy, Fernandes:2019zuz, Felkai:2022cmb}, and via the application of Richardson-Lucy deconvolution~\cite{RLdecon}. 
Application of either classical or machine-learning based approaches to the recorded events can then be used to make a powerful single- vs double-electron discrimination cut~\cite{Ferrario:2019kwg, Kekic_2021}. 

The NEXT program has completed 2 demonstrator-scale phases, with prototypes \\ \noindent {\tt NEXT-DBDM} \cite{Alvarez:2012yxw} and {\tt NEXT-DEMO}~\cite{Ferrario:2015kta} providing the first proofs of energy resolution and topological event identification respectively, followed by {\tt NEXT-White}~\cite{Monrabal:2018xlr} , the first underground and radio-pure phase of the program. 
{\tt NEXT-White} has demonstrated energy resolution of better than 1\% FWHM at Q$_{\beta\beta}$, validated the performance of the topological discrimination~\cite{Ferrario:2019kwg} as well as the NEXT background model~\cite{Novella:2019cne}, and measured the $2\nu\beta\beta$ half-life via direct background subtraction between enriched and natural xenon gas~\cite{Novella_2022}, the first experimental proof of this technique. 
The {\tt {NEXT-100}} experiment~\cite{Alvarez:2012as} is now being constructed, which will contain 100~kg of enriched xenon gas underground at the Laboratorio Subterr\'{a}neo de Canfranc (LSC) with a sensitivity to $0\nu\beta\beta$ of $2.8\times10^{25}$ years with three years of operation. 
Ton-scale conventional phases of this program are under development~\cite{adams2021sensitivity} with a sensitivity of $1.4\times10^{27}$ years after 5~years of operation, as well as an active R\&D program to enable detection of the Ba$^{2+}$ daughter ions emitted in the decay through single molecule fluorescence imaging~\cite{jones2016single,mcdonald2018demonstration,byrnes2019barium,thapa2019barium,thapa2021demonstration,Rivilla:2019vzd,herrero2022ba,jones2022dynamics}.

In the current NEXT experimental designs, ionization electrons are driven by a high negative voltage cathode toward an EL region near zero volts (ground), and the event topology is measured with a plane of SiPMs located behind a wavelength shifting, TPB-coated plane. 
Positive ions, including the daughter ion of the decay Ba$^{2+}$, drift toward the negative high voltage cathode. 
In this work, we consider an alternative solution that replaces the function of the SiPM tracking plane with detection of secondary scintillation light at a large distance from a high positive voltage EL plane. Fast optical cameras coupled with VUV-sensitive image intensifiers enable direct observation of the $\sim$172~nm light from the distant EL region.  
This concept builds on past work on camera-based time projection chambers by others, in particular in two-phase liquid argon TPCs such as ARIADNE~\cite{ARIADNE,Roberts_2019, lowe2023ariadne}, directional dark matter detectors with a CCD readout~\cite{ahlen2011first, phan2016gem}, and imaging techniques for nuclear decays~\cite{miernik2007optical}.

For the $0\nu\beta\beta$ application, camera-based readout may confer important benefits if the tracking performance can be proven competitive with existing readout techniques. 
Within this new paradigm, all electronic elements of a large tracking system would reside outside the detector volume, with positive implications for the management of system complexity, radiopurity and cost. 
These issues become especially pressing as the tracking region becomes larger and more granular, as envisaged within future large-scale experiments~\cite{adams2021sensitivity}.  
In this scheme, the cathode is almost fully unoccupied by sensors, a potentially key benefit for the future realization of a barium tagging system, which would hope to use as much of the cathode area as possible for collecting barium ions.  
Moreover, the fact that only VUV light is detected allows the removal of internal organic wavelength shifters from the volume, of benefit for simplifying the understanding of the optical behaviour of the device, but also crucial if a low-background single-molecule sensitive barium detection scheme using single molecule fluorescence imaging is to be realized.
Using mirrors to reflect the image, the external camera and transfer optics also need not align directly with the view port through which it observes the track topology.
This means that the camera position can be easily controlled, allowing for shielding to be placed between the camera optics and the vessel, reducing concerns that the camera electronics might introduce background radiation.
If suitable performance can be proven, this solution would enable a highly scalable implementation of a large-area EL tracking detector that naturally admits integration of a barium tagging cathode. 
This concept, called Camera Readout and Barium Tagging (CRAB)~\cite{adams2022camera}, is the subject of an ongoing R\&D program at the University of Texas at Arlington and Argonne National Laboratory within the NEXT collaboration.  
The {\tt NEXT-CRAB-0} detector described here is the first in a sequence of CRAB demonstrators that aim to prove this technology for ton- to multi-ton-scale application.

\begin{figure}[h!]
\centering
\includegraphics[width=0.99\linewidth]{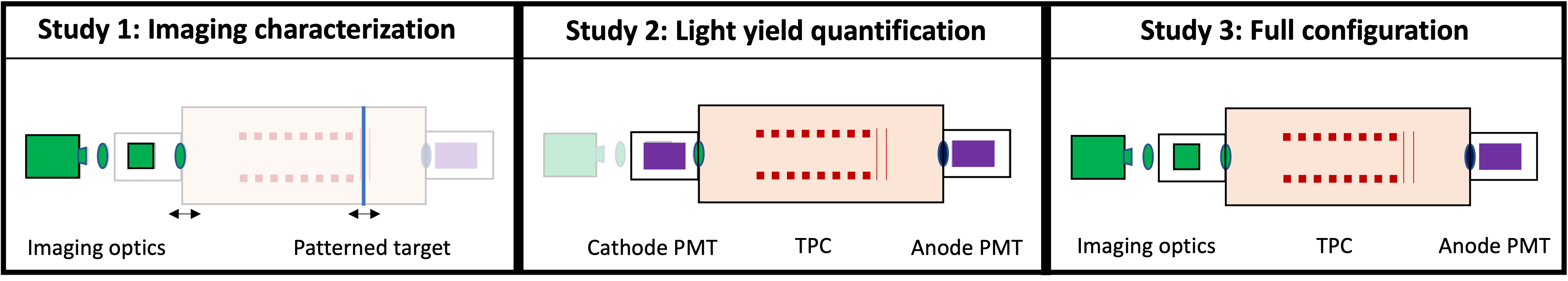}
\caption{The three system configurations in which experimental data reported in this paper were collected.}
\label{fig:ThreeStudies}
\end{figure}

This paper presents a detailed technical description and first results of the {\tt NEXT-CRAB-0} system.  
In Sec.~\ref{sec:crab-0}, we describe the detector design with details of the electronics, gas system,  optical components and calibration source.  
Section~\ref{sec:simulations} describes {\tt GEANT4}-based simulations and optical calculations against which the data are bench-marked.  
Section~\ref{sec:ExperimentalResults} then presents the results of experimental studies made in three system configurations, shown schematically in Fig.~\ref{fig:ThreeStudies}.  
First, the imaging system is characterized {\em ex-situ} in order to establish the proper protocol for focusing and alignment (Sec.~\ref{sec:OpticsTests}).  
Second, the light yield from the TPC is measured and compared against simulations using PMTs at both windows to calibrate brightness of the camera-based intensified images (Sec.~\ref{sec:light-yield}). Finally, the system is operated in its full configuration, delivering first tracks from alpha and beta rays as well as from cosmic ray muons (Sec.~\ref{sec:signal-proc}).  
The paper concludes with a discussion of the next steps in this research program and the implications for the barium tagging program in NEXT (Sec.~\ref{sec:Conclusion}).

\FloatBarrier
\section{The {\tt NEXT-CRAB-0} detector}
\label{sec:crab-0}
The Camera Readout and Barium Tagging V.0 ({\tt NEXT-CRAB-0}) is a pathfinder detector that aims to take the first steps toward demonstrating VUV image intensified readout of a high pressure xenon gas TPC. 
This publication presents the first results of {\tt NEXT-CRAB-0}, whose goals are to validate the optical-system designs and overall detector configuration by proving high resolution transverse imaging of tracks from the 5.3~MeV alpha particles from Po-210 decays and beta electrons from $^{210}$Bi with Q-value 1.16~MeV.

\subsection{TPC design}

\begin{figure}[t]
\centering
\includegraphics[width=0.99\linewidth]{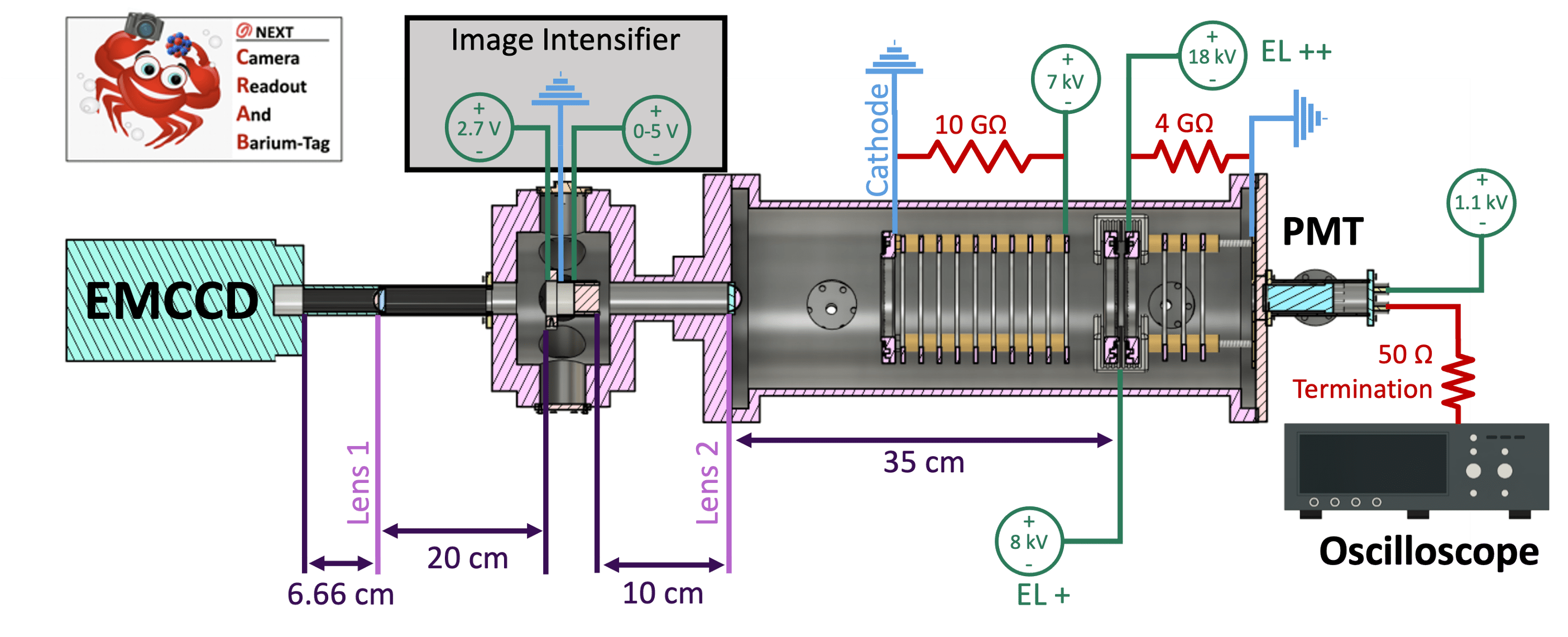}
\caption{{\tt NEXT-CRAB-0} detector in outline.  
Key dimensions and electrical connections are shown.}
\label{fig:CRAB-0-connections}
\end{figure}

{\tt NEXT-CRAB-0} is an asymmetric xenon gas TPC using EL with three sections, defined axially from left to right in reference to Fig.~\ref{fig:CRAB-0-connections}.   
The largest is the drift region,  where the ionization electrons are transported in a drift electric field towards the EL region.  
The  electric field uniformity is maintained by a field cage (FC), whose electrostatic boundary conditions are defined using a field cage created by stacking aluminum rings on PEEK rods and separating them with 1/2 inch ceramic beads. 
The rings are vented with grooves on each of the three mounting holes to avoid virtual leaks. 
The potential is graded with a chain of 1~G$\Omega$ resistors mounted inside the field cage between every pair of rings. 
The drift field is created by electrically biasing two ends of the cage.  
The cathode is situated at one end, connected directly to detector ground; at the other end is the high voltage EL region, followed by a buffer region that steps the anode voltage down over three further field grading rings.

The EL region is delineated by two photo-etched meshes with 7~mm separation.  
EL regions in HPGXe TPCs must meet important design criteria in order to achieve the superb energy resolution and uniform topological reconstruction that is accessible using the EL amplification technique.  
These include  mechanical flatness of the EL region under electrostatic deflecting forces,  high photon transparency of the EL region and good field uniformity in the EL gap.  
Those conditions must be met while remaining stable against electrostatic discharge and robust against damage in the event of sparking.

\begin{figure}[t]
    \includegraphics[height=0.4\linewidth]{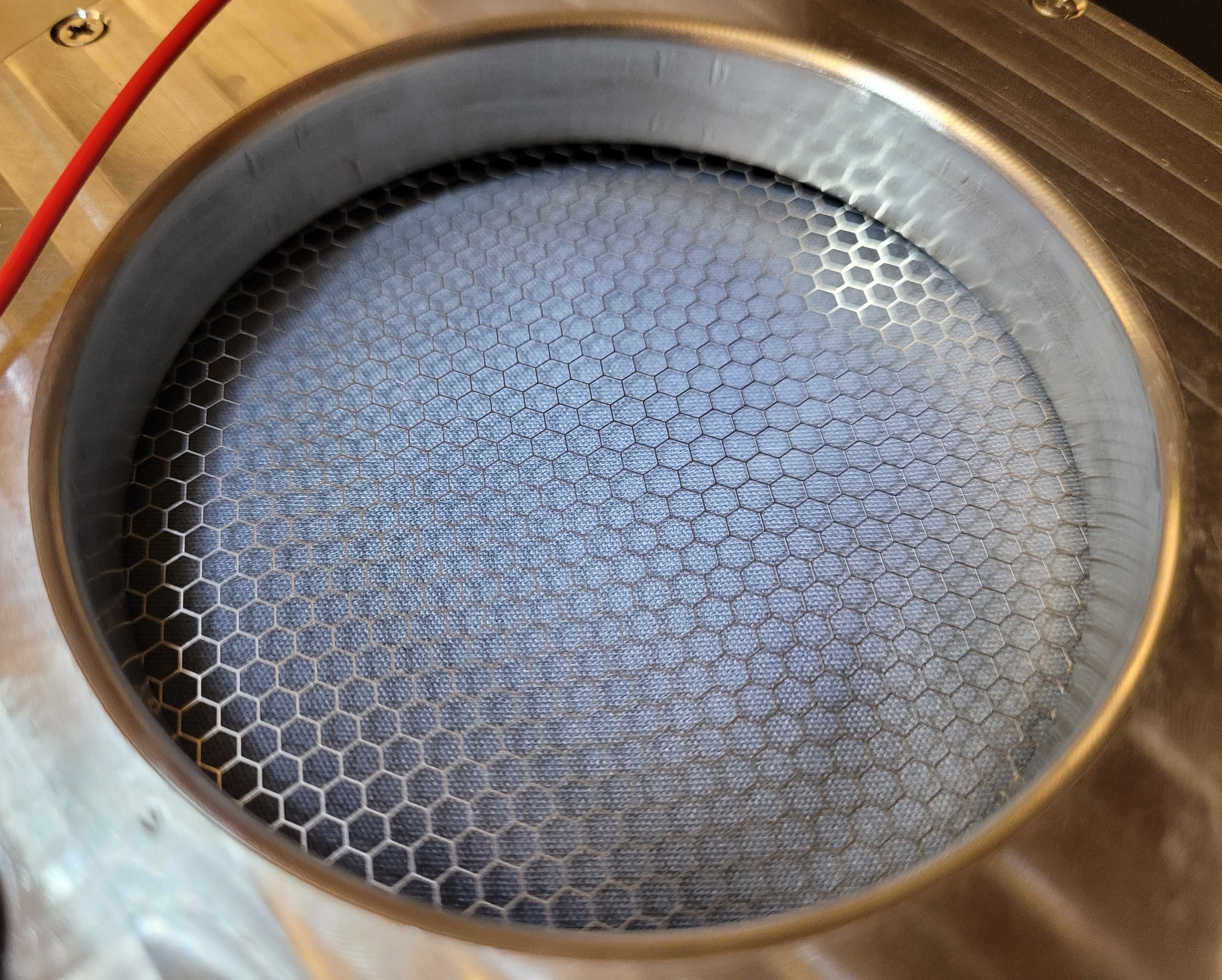}
    \includegraphics[height=.4\textwidth]{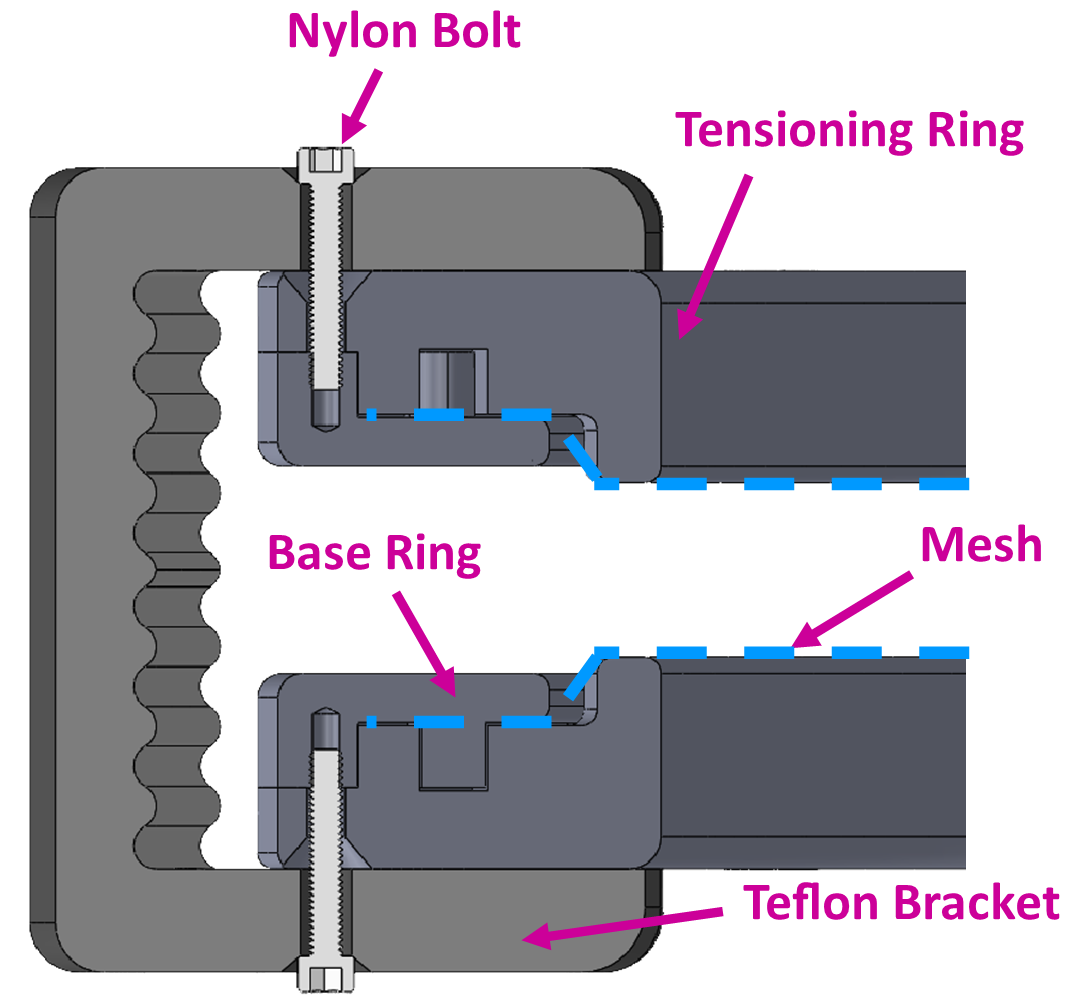}
    \caption{Left: Picture of mesh tensioned within its frame. The frame that holds the mesh is also used to do the tensioning.  
    Right: Section view of the EL region. As the two frames are bolted together, the lip of the top ring presses down on the mesh and stretches it past the base ring as shown by the blue dashed line. 
    The nylon bolts replace some of the metal-tensioning screws to hold the rings in place on the Teflon brackets.}
    \label{Fig:tensioningmethod}
\end{figure}

Although the {\tt NEXT-CRAB-0} detector has a small EL region of inner diameter 7.2~cm, the EL frames and mesh designs are scale models of parts for the forthcoming {\tt NEXT-100} experiment, which will have an imaging region of 1~m inner diameter~\cite{Alvarez:2012haa}.  
A 90\% transparent hexagonal mesh is photochemically etched from 75 micron-thick stainless steel with a honeycomb pattern that maximizes the strength per mass of material and avoids the saddle-like deformation mode of larger frames under higher tensions, problematic with square meshes.  
The mesh is tensioned using its outer solid rim between two stainless steel rings, as shown in Fig.~\ref{Fig:tensioningmethod}.  
Three Teflon brackets hold the meshes parallel and in place, as shown in Fig.~\ref{Fig:tensioningmethod}, right, screwed into the mesh frames using insulating bolts.  
Ridges along the bracket surface inhibit electrical breakdowns and movement of surface charges at the EL boundary. 
The buffer further aids in the prevention of breakdowns, stepping down the voltage of the higher potential EL ring to ground using four 1 G$\Omega$ resistors connecting, in series, the EL ring, the three field grading rings, and the vessel wall.
The EL region and buffer are mounted onto the pressure vessel endcap.
The cathode is formed from a similarly tensioned photo-etched mesh and is electrically connected to the top of the field cage, both mounted into the pressure vessel body via a machined aluminum bracket.

\subsection{Detector high voltage}

High voltage is supplied to internal field cage elements via epoxy-potted coaxial RG-59 cable feedthroughs. Inside the detector, each cable is terminated at a lug connector coupled to the relevant electrode by a screw connection.  
External connections are made within high-voltage splice boxes to three Glassman Series FJ power supplies. 
All supplied high voltages are positive in polarity, the highest being supplied at the EL anode (EL++), followed by the EL gate (EL+), both penetrating through the vessel endcap. 
The first field cage ring is connected with a cable penetrating through the pressure vessel radially, with the cathode (C) is held at ground by an electrical connection directly to the pressure vessel wall. 
Typical high voltage operating parameters are shown in Table \ref{table:MagOfEFields}. 
The errors on voltages come from voltage accuracy of power supplies that are obtained from the manufacturer's manual \cite{PowerSupply} while the errors on lengths come from measured values, and these errors are propagated to obtain the errors on the electrical fields.

The image intensifier power and gain settings are controlled via a BK Precision 9132B power supply, with three cables penetrating the auxiliary vacuum vessel through a multi-pin commercial feedthrough.  
The image intensifier settings and detector high voltage are all connected via custom {\tt LabVIEW} slow controls system, described in Appendix A.

\begin{table}[!ht]
\begin{center}
\centering
\begin{tabular}{|p{2cm}||p{3cm}||p{3cm}| |p{2cm}||p{3cm}| }
 \hline
 Region & Boundary & $\Delta${V} (V) & L (cm) & E (V/cm) \\ 
 \hline
 Drift &C to FC & $7.04\times10^{3}  \pm 40 $ & $16.0 \pm 0.1$ & $4.40\times10^{2} \pm 4$ \\
 Interface &FC to EL+ & $9.90\times10^{2}  \pm 60 $ & $2.1 \pm 0.1$ & $4.71\times10^{2}  \pm 36$ \\ 
 EL &EL+ to EL++ & $9.98\times10^{3}  \pm 110 $ & $0.71 \pm 0.03$  & $1.40\times10^{4} \pm 600$ \\
 \hline
\end{tabular}
\end{center}
\caption{Applied electric fields for the data sets presented in this paper. $\Delta${V}  is the total applied voltage, $L$ is the length of the region, and $E$ is the electric field. All data were collected at 9.7~bar of Xe gas.}
\label{table:MagOfEFields}
\end{table}

\begin{figure}[t]
\centering
\includegraphics[width=0.27\linewidth]{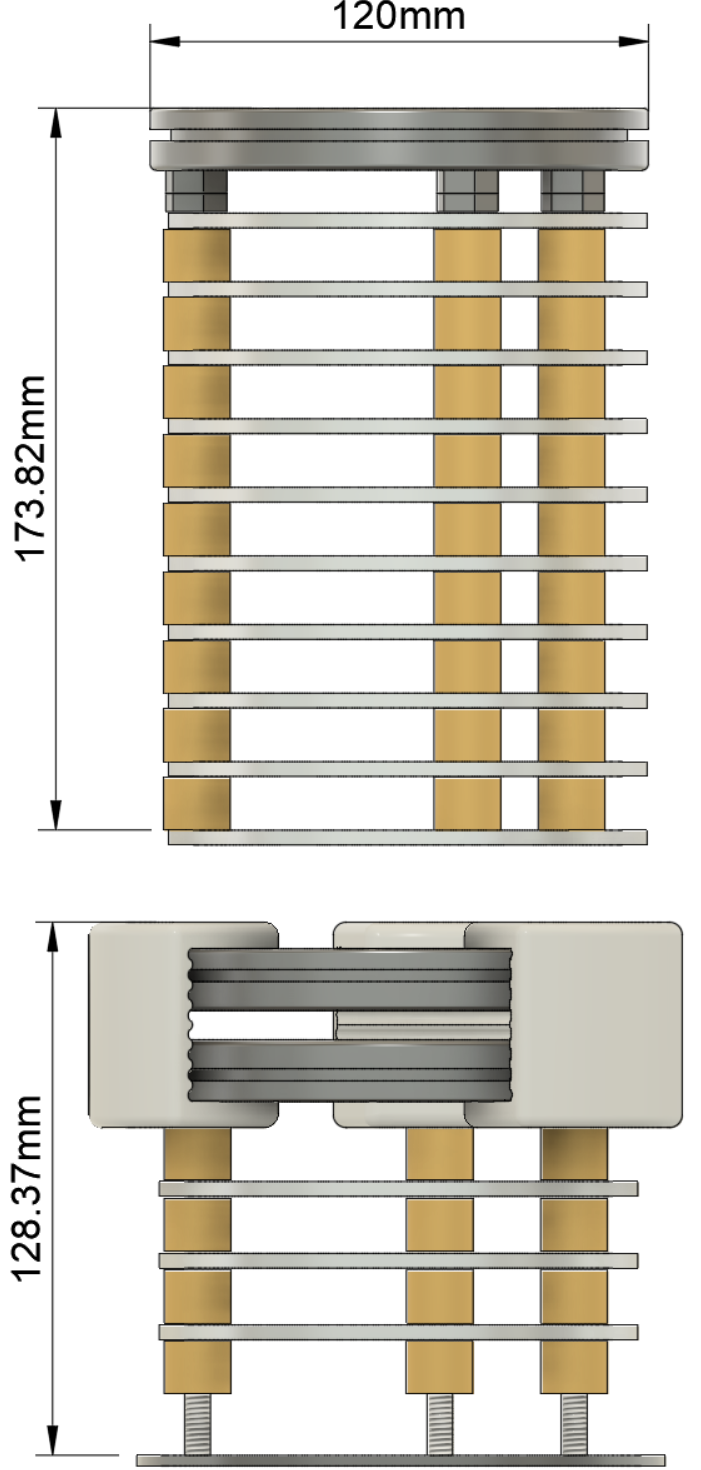}
\includegraphics[width=0.46\linewidth]{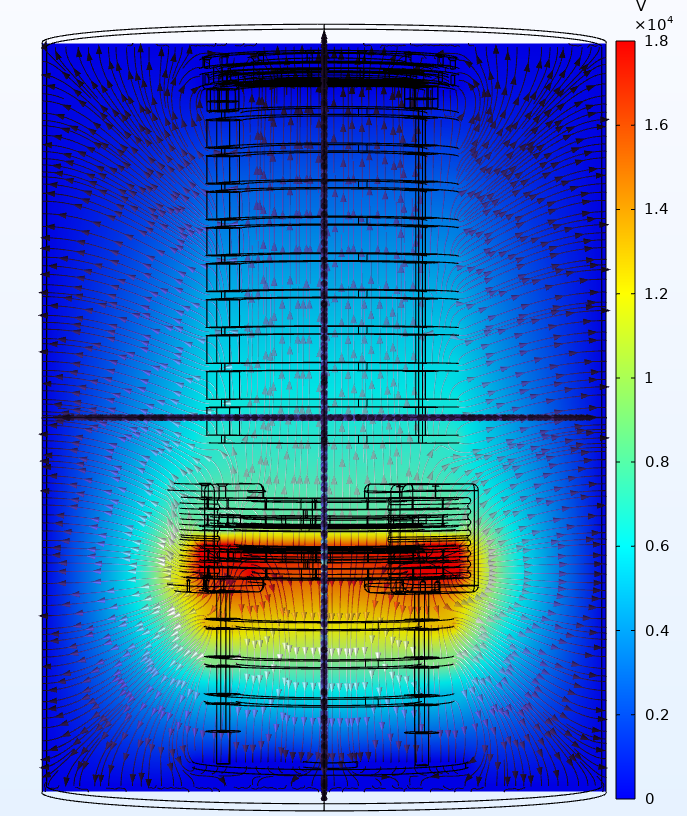}  
\caption{Left: A schematic of the inner elements of the TPC. 
Right: An electric potential simulation of the inside of the chamber using {\tt COMSOL}. 
The colors in the COMSOL simulation represent the electrical potentials applied while the lines  represent the electric field, with the arrows pointing in the field line directions.}
\label{fig:OpTPCEFields}
\end{figure}

A full internal field map was simulated using the {\tt COMSOL} Multi-Physics simulation tool~\cite{dickinson2014comsol}. The 3D geometry was drawn in the Fusion 360 CAD software and then imported into {\tt COMSOL}, with voltages as specified in Table~\ref{table:MagOfEFields}.  
The optical TPC is simulated in a stainless steel chamber with xenon gas as the background medium and material properties associated with each component in the TPC, including the polarizable dielectric elements.  
The results of this simulation are shown in Fig.~ \ref{fig:OpTPCEFields}, right.  
The simulation verifies both that near-perfect electron transmission into the EL region is expected, and also that there are not expected to be any problematic regions of high field due to fringing effects. The highest field in the geometry occurs between the EL meshes and is consistent with expectations for a purely parallel plate geometry.

\subsection{Gas handling}
\label{sec:Gas System}

\begin{figure}[t]
    \centering
    \includegraphics[width=0.99\linewidth]{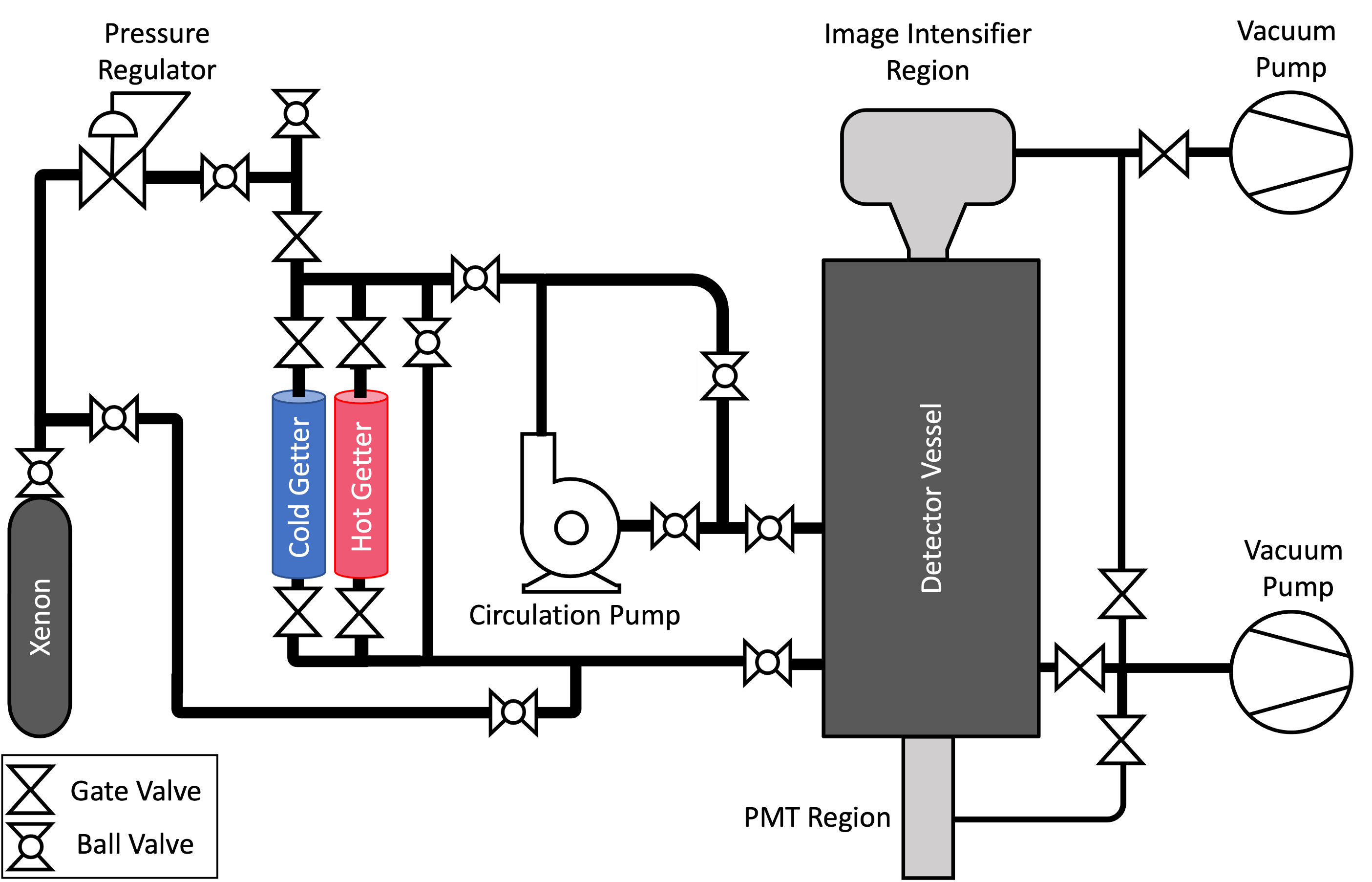}
    \caption{Diagram of the gas gystem used for \tt{NEXT-CRAB-0}.}
    \label{fig:gas_sys}
\end{figure}

The TPC is supplied with purified, pressurized xenon gas that is circulated through cold (SAES MicroTorr HP190-902F) and hot (SAES MonoTorr PS4-MT3-R-1) getters to remove water, oxygen and nitrogen contamination to part-per-billion levels of impurity.
This filtering process is done in two steps, first cleaning with the cold getter to remove most reactive gas species, followed by cleaning with the hot getter to remove any remaining gas contaminants.
Xenon from a gas cylinder is used to pressurize the vessel to $\sim$10~bar,  monitored by an electronic pressure gauge installed on the vessel.  
Before filling, the vessel is evacuated to $\sim 1\times10^{-6}$ mbar by a Pfeiffer HiCube 80 Classic Turbopumping station, followed by recirculation and cleaning of the gas for several hours to remove residual water and oxygen. 
A separate Pfeiffer HiCube 80 Eco is used to evacuate the image intensifier and PMT chambers.   
Between runs, the xenon gas is collected via cryogenic condensation using liquid nitrogen.   
A schematic diagram of the gas system is shown in Fig~\ref{fig:gas_sys}.

\subsection{Optics}

All light in the {\tt NEXT-CRAB-0} system is collected  directly in the VUV, with no internal wavelength shifting applied.  
Both ends of the detector are instrumented with VUV light collection systems which sit in vacuum chambers that are separated from the detector volume by custom made magnesium fluoride (MgF$_2$) pressure windows.  

At the EL end, a single Hamamatsu R7378A PMT is used for triggering and studies of event timing.  
In order to detect UV light the PMT is housed within an auxiliary vacuum chamber separated from the gas volume by a bespoke MgF$_2$ planar pressure window. The PMT signal is recorded using a Teledyne Waverunner 8054 digital oscilloscope, terminated at 50 $\Omega$ and sampling at 5 giga-samples/s.

\begin{figure}[t]
    \centering
    \includegraphics[height=0.45\linewidth]{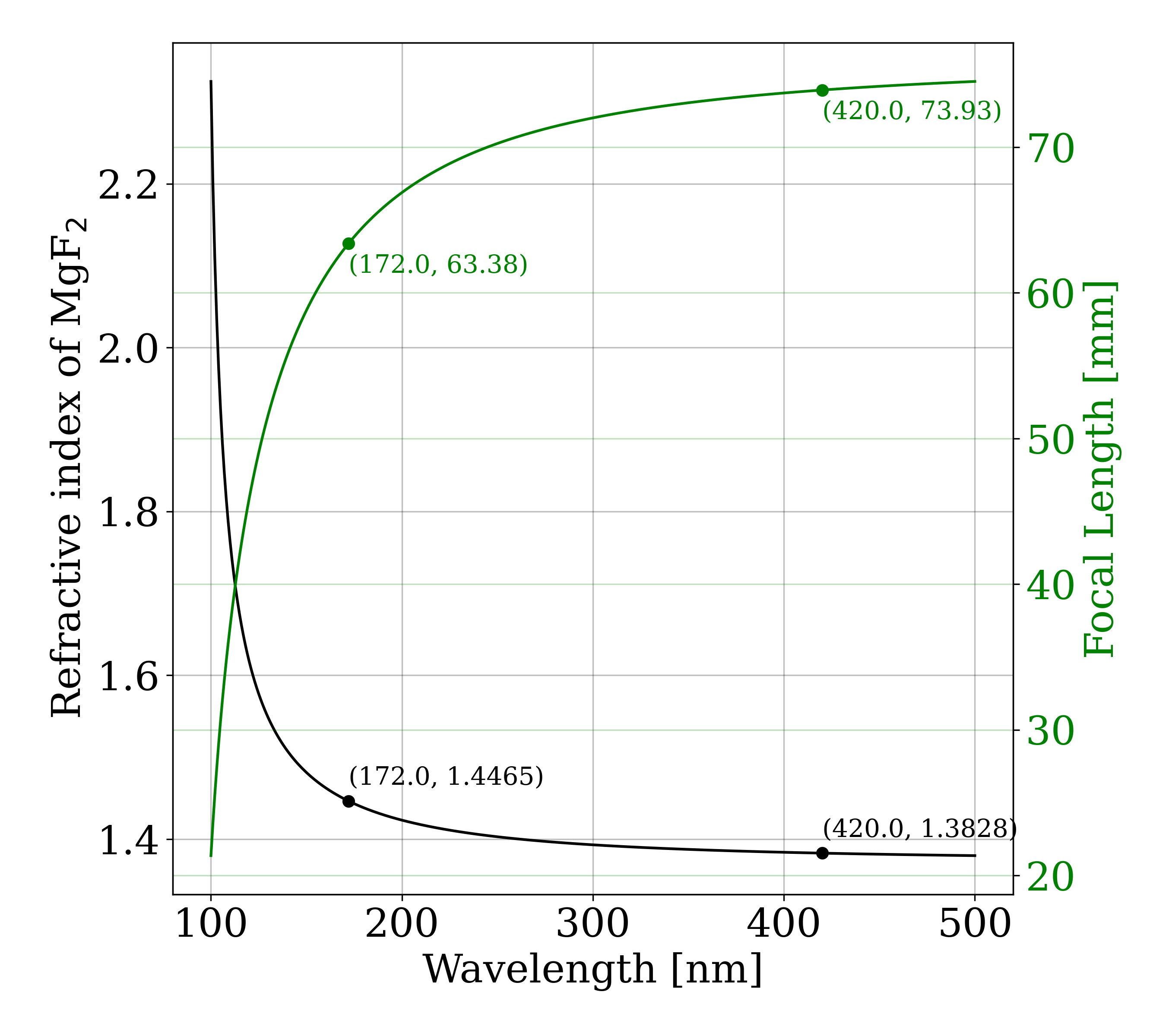}
    \caption{Refractive index and focal length of a 28.3~mm radius of curvature MgF$_2$ lens vs wavelength.}
    \label{fig:MgF2}
\end{figure}

At the cathode end sits the tracking system, comprised of the VUV focusing lens and image intensifier housed within a second auxiliary vacuum chamber, viewed via transfer optics by external camera.  
172~nm light generated in the EL region of the detector first encounters the MgF$_2$ pressure window which has a 16~mm transparent diameter. 
This window is plano-convex with a radius of curvature of 28.3~mm, focusing light from the EL onto the photo-cathode of a  VUV-compatible image intensifier.  
The lens and image intensifier must be aligned before sealing the vacuum chamber and without the benefit of UV light for tuning.  
This is accomplished by extrapolating the focal positions established with visible light in the studies of Sec.~\ref{sec:OpticsTests} to 172~nm using the Sellmeier Equation~\cite{andp.18722231105} to treat the refractive index evolution of MgF$_2$ to shorter wavelengths (Fig.~\ref{fig:MgF2}).  
The resulting focal length  at 172~nm is 63~mm.  
Internally, the image intensifier converts the VUV light arriving at its photocathode into photoelectrons that are amplified by a microchannel plate (MCP) inside the device, with gain tunable by setting the MCP voltage, with a maximum specification of approximately 3,000$\times$.  
The image intensifier in {\tt NEXT-CRAB-0} was custom-made for our application by Photonis with a quantum efficiency of ~22\% at 172 nm.  
The amplified charge strikes a phosphor plane, emitting light peaking at 530 nm, which is focused by a BK7 planoconvex  glass lens (lens 1 in Fig.~\ref{fig:CRAB-0-connections}) onto the camera CCD via a glass vacuum view port.

The camera used in these studies is a Hamamatsu {\tt ImagEM X2} Electron Multiplying CCD which has a 1200$\times$ photon gain factor, allowing for single photon imaging at the camera interface, as well as a 90\% quantum efficiency .  
The data is read out over an IEEE 1394 conduit to a PC running commercially available camera readout software, which stores the images in TIFF format for further processing. 
While this particular camera is not sufficiently fast to enable full 3D readout, the collection of two-dimensional projections of the track  provides a key proof-of-principle for the CRAB approach.  
A full 3D demonstration will follow in a subsequent phase of the program, as discussed in Sec.~\ref{sec:Conclusion}. 

\begin{figure}[t]
    \centering
    \includegraphics[height=0.45\linewidth]{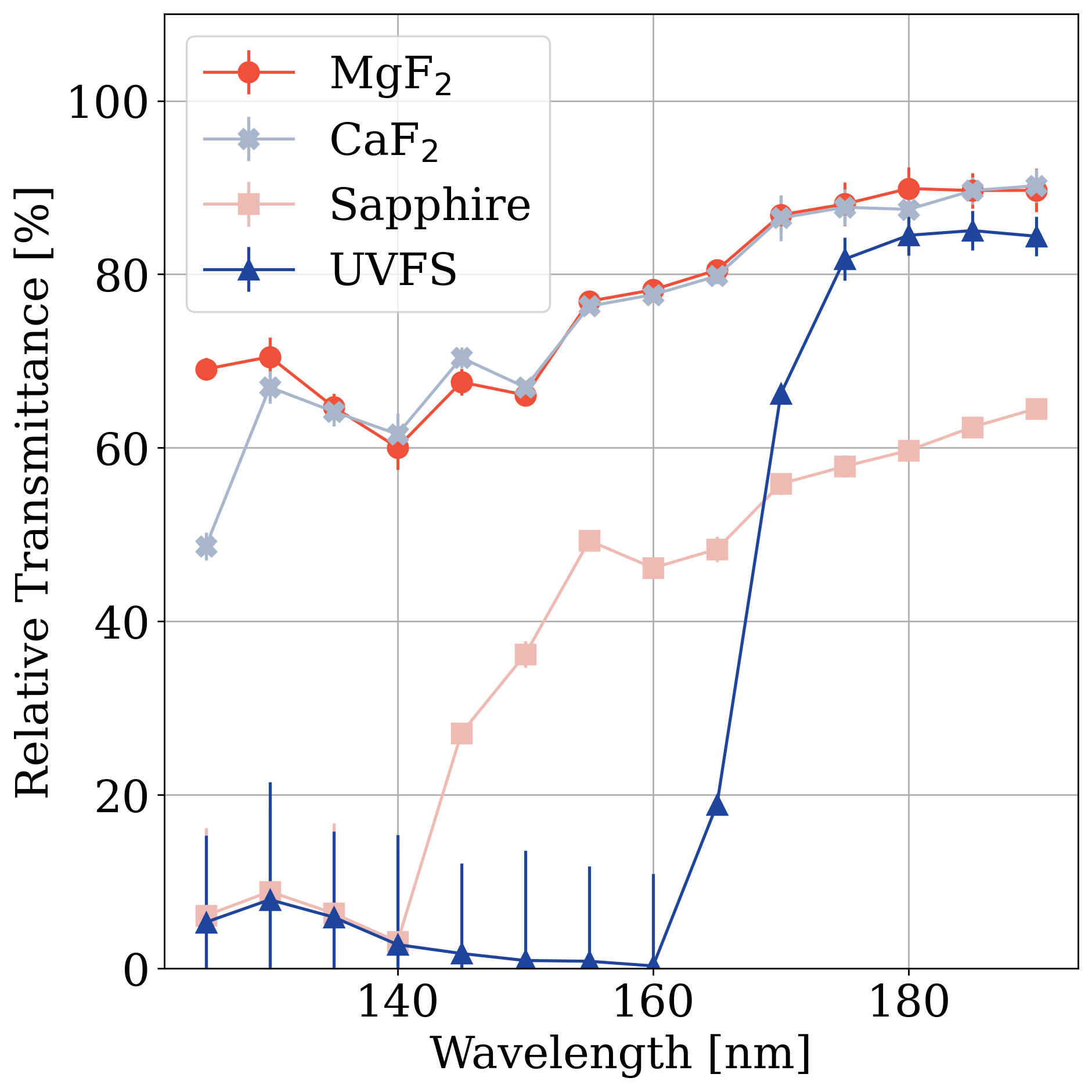}
    \includegraphics[height=0.45\linewidth]{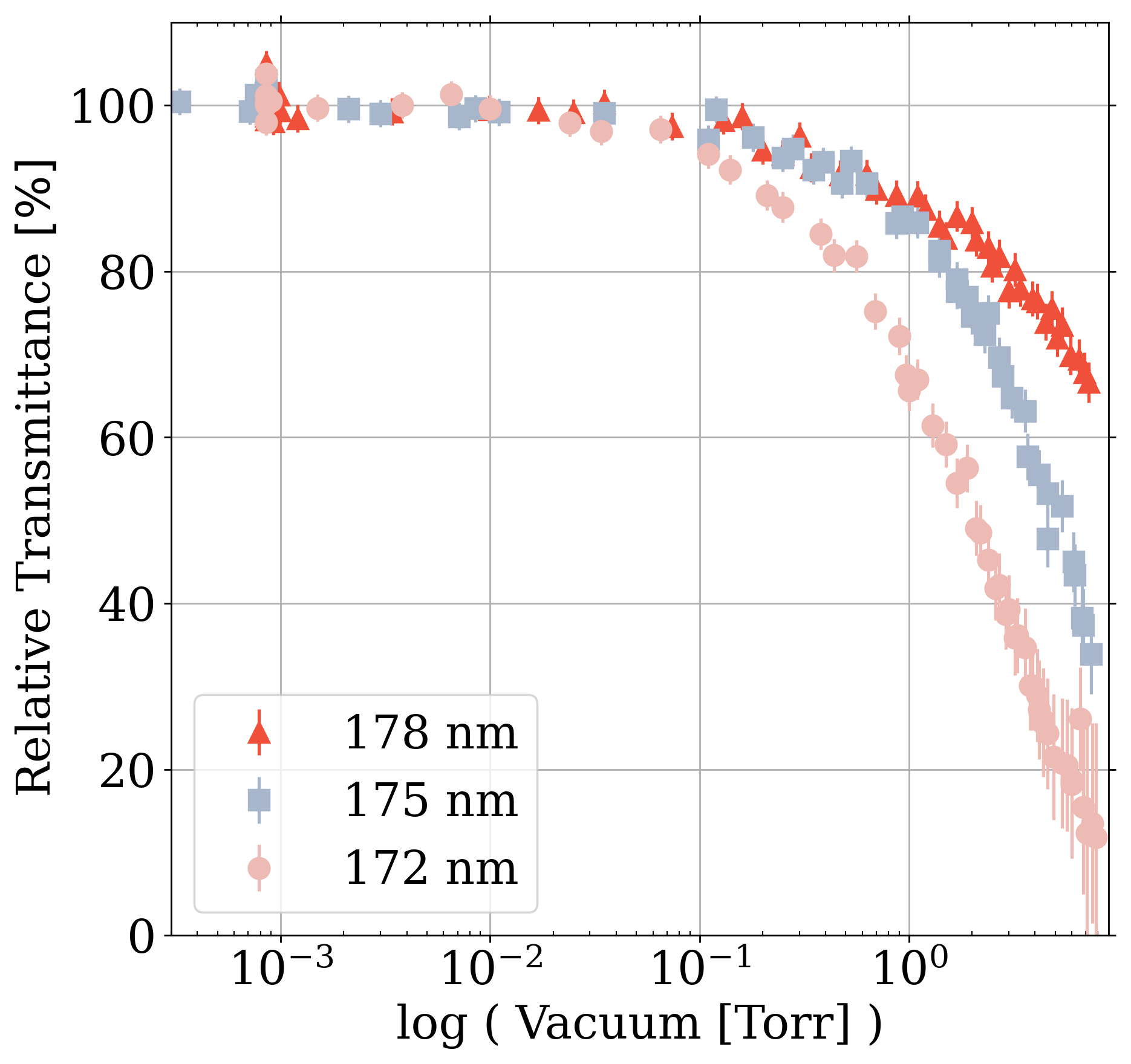}
    \caption{Measurement of relative transmittance for 5~mm-thick windows of various materials (Left) and vacuum levels (Right), as measured with a deuterium lamp and diffraction grating to select wavelength. 
    Uncertainties in measurements are statistical only. }
    \label{fig:transmissivity}
\end{figure}

Preliminary experiments were undertaken using a VUV monochromator to select an appropriate window lens and to establish the vacuum quality necessary for the  auxiliary vacuum chambers that house the image intensifier and PMTs (Fig.~\ref{fig:transmissivity}).  
A vacuum vessel was coupled to a deuterium lamp with a broad emission spectrum, and a diffraction grating with a resolution of 1~nm selects specific wavelengths to measure transmissivities. 
A UV-sensitive PMT is placed in the optical path within the vacuum chamber to collect the diffracted light. 
Transmissivities are measured by taking ratios of the light observed with optical elements in the PMT path to those in vacuum.   
Fig.~\ref{fig:transmissivity} upper left shows the transmissivity of various windows tested for the CRAB program.  
MgF$_2$ windows were found to have high transmissivity at 172~nm with an absolute value of around 87$\pm$2\%, and good mechanical properties for holding pressure. 

To establish the vacuum requirements for the auxiliary vessels, a comparison of transmittance through different vacuum qualities was made, and shown in Fig.~\ref{fig:transmissivity} upper right.  
The total absorption of VUV light through air depends on wavelength, pressure, and total path length.  
For this measurement, the length of our diffraction grating test was approximately 750~mm, which is long compared to the optics region {\tt NEXT-CRAB-0}, and so good transmissivity here setup provides a conservative upper limit for the required vacuum level.  
Results show that a modest vacuum quality of $\leq$0.1 Torr is appropriate to ensure transmission of at least 99\% of the VUV scintillation light. 
This is comfortably achievable in both auxiliary vacuum housings, which operated at a vacuum level of 10$^{-5}$ Torr for the experiments described in this paper.

\subsection{Calibration source}
\label{sec:calibration}

A $^{210}$Pb needle source is installed in the detector to allow for calibration. 
The $^{210}$Pb needle decays via low-energy beta to $^{210}$Bi. 
This daughter nuclide subsequently beta decays, emitting a 1.16~MeV beta electron  which is partially contained and gives a useful tracking candidate, producing a further daughter $^{210}$Po. 
The $^{210}$Po then decays via alpha emission with 5.3 MeV of energy, producing an alpha particle which has a range of $\sim$2~mm in 10 bar xenon gas. 
This alpha provides a high energy calibration peak. The $^{210}$Pb needle source is installed in the drift region, mounted on the field cage approximately 5~cm up-drift of the EL Region.  
The source activity of each process can be inferred from the observed $\alpha$ decay rate on the PMT to be approximately 12.5~Bq, which corresponds to an average of 0.35 alpha particles and 0.35 beta particles per camera frame.  
With this source rate we expect 22.8\% of events to contain one alpha track and no other activity, 22.8\% to contain one beta track and no other activity, 42.2\% empty, and the remaining 12.2\% to have some degree of pile-up with multiple tracks.  
The PMT waveforms, on the other hand, are sufficiently fast that the pileup rate in triggered pulses is negligible.

\section{Simulations}
\label{sec:simulations}
The {\tt NEXT-CRAB-0} detector has been fully simulated in {\tt NEXUS}, the {\tt GEANT4}-based software package developed for Monte Carlo simulations of NEXT detectors \cite{NEXUS}. 

\subsection{NEXUS {\tt GEANT4} simulations}
\label{sec:Light Yield Simulations}

\begin{figure}[t]
\centering
\includegraphics[width=0.99\linewidth]{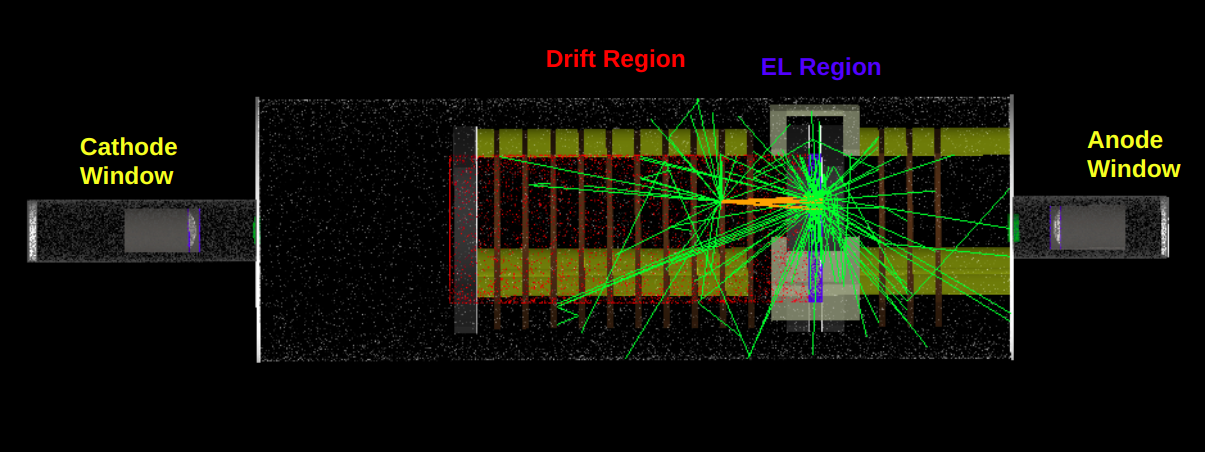}
\caption{The {\tt NEXT-CRAB-0} detector geometry rendered in {\tt GEANT4}.  }
\label{fig:Sim_full}
\end{figure}
To simulate the {\tt NEXT-CRAB-0} detector in {\tt NEXUS}, a cylindrical stainless steel geometry was created containing a MgF$_2$ lens with a thickness of 5~mm and diameter of 1.66~cm at the cathode side while a MgF$_2$ window with a thickness of 6~mm and diameter of 1.62~cm was placed at the anode side. 
The simulation also contains the stainless steel EL meshes, mesh rings, and all relevant field-cage components and materials. 
Figure ~\ref{fig:Sim_full} illustrates the visualization of the {\tt NEXT-CRAB-0} detector in GEANT4.
The MgF$_2$ windows refractive index is parameterized using coefficients from Ref.~\cite{andp.18722231105}.

Pure xenon gas with a pressure of 9.7 bar is simulated as the active medium. 
{\tt NEXUS} internal routines are used to model clustering, drift, and EL. 
Clustering converts each deposited energy by a charged particle to a number of ionization electrons, and drift models the diffusive and directed motion of ionization electrons under a uniform electric field.  
The transport parameters including transverse and longitudinal diffusion constants and drift velocities are calculated using the {\tt PyBoltz} transport code for each region~\cite{Pyboltz}.  
These values are listed in Table~\ref{table:Diffusion_Constants}.

\begin{table}[b!]
\begin{center}
\centering
\begin{tabular}{ |p{3cm}||p{3cm}||p{3cm}||p{3cm}|   }
 \hline
 Region &$v_D$ $(\frac{\mathrm{mm}}{{\mathrm{\mu s}}})$& D$_T(\frac{\mathrm{mm}}{\sqrt{\mathrm{cm}}})$& D$_L(\frac{\mathrm{mm}}{\sqrt{\mathrm{cm}}})$ \\
 \hline
 Cathode To EL+ & 0.90 $\pm {0.04}$ & 0.92 $\pm{0.05}$ & 0.36 $\pm{0.01}$ \\
EL+ to EL++ & 4.6 $\pm{0.2}$ & 0.24 $\pm{0.01}$ & 0.17 $\pm{0.01}$ \\
 \hline
\end{tabular}
\end{center}
\caption{Drift velocity, longitudinal and transverse diffusion constants from {\tt PyBoltz} used in the {\tt NEXUS GEANT4} simulation of {\tt NEXT-CRAB-0.}} \label{table:Diffusion_Constants}
\end{table}

The ionization and scintillation yields are parameterized based on the work of ~\cite{Smearing}, with 45,455 electrons / MeV and 25,510 photons / MeV respectively. 
Secondary scintillation light is generated in the EL region and produces isotropic photons randomly sampled along the electron track.  
The EL region has a 0.7~cm gap in which each electron produces secondary scintillation with a gain of 925 photons per cm per electron, using the reference yield from \cite{monteiro2011determination}.  
For the $\alpha$-particles a Gaussian distribution with width 3\% for S2 was convolved with the final spectrum to account for random charge sharing through recombination, following the measurements of \cite{Smearing}. 

To account for the effects of partial occlusion of both charge and light in the immediate vicinity of the needle source, a simplified needle geometry is encoded, as a solid cylinder with a circular eye.  
All decay daughters are injected 2~nm away from the needle surface, and electrons or photons touching the needle volume are assumed to be absorbed.

To calculate the expected S2 light yields, simulated PMTs are positioned at the MgF$_{2}$ window apertures, each enclosed by a cylindrical tube under vacuum.  
Photons reaching each photocathode are recorded and counted to determine the detected light per event.   
The absolute quantum efficiency of each PMT is not precisely known and is especially difficult to characterize accurately at 172~nm. 
The Hamamatsu data-sheet gives a typical value of 13.9\%, though suggests variability in cathode radiant sensitivity between devices of around $\pm$30\%.   
This is the dominant systematic uncertainty in the absolute light yield prediction, and we find that a 9.7\% quantum efficiency provides a quantitatively good match to data for both anode and cathode PMTs.

In our primary calculation of expected light yield, each VUV photon is simulated through the detector geometry using {\tt GEANT4}. 
Stainless steel is parameterized with a reflectivity of 20\%, Rayleigh scattering behaviour in xenon is incorporated, and photons are deflected at angles predicted by classical electromagnetism at xenon-MgF$_2$ and MgF$_2$-vacuum interfaces. 
Transparency of the meshes has been taken into account, and final results are scaled by the transmission efficiency of the windows.
Finally, photons striking PMT photocathodes are randomly down-sampled based on the detector quantum efficiency.  

At least two different central wavelengths of scintillation photons are reported in the literature at 10~bar xenon, 170~nm and 172~nm. 
According to digitized values of quantum efficiency from the PMT manual~\cite{PMT}, there is 0.6\% difference in quantum efficiency and also there is 1~\% difference between the measured and extrapolated MgF$_{2}$ transmission efficiencies of 170~nm and 172~nm~ Fig.~\ref{fig:transmissivity}. 
We have simulated separate 5.3~MeV alpha events in 9.7~bar xenon with the xenon electroluminesence emission spectrum peaking at 170~nm and 172~nm, obtaining a similar number of photoelectrons in both windows by applying corresponding quantum efficiencies and MgF$_{2}$ transmission efficiencies for each wavelength. 
Results indicate that there is a 1.8$\pm{0.1}$\% difference at the anode window and a 1.4$\pm{0.1}$\% difference at the cathode window, which is negligible compared to other sources of systematic uncertainty.
This photon tracking simulation is used to  determine the PMT quantum efficiency from alpha particle data, as well as to benchmark the faster geometric optics based approach used to simulate camera images.

In the geometric optics approach, each ionization electron is assumed to be an isotropic emitter of photons in the EL region with a fixed mean yield.  
This number is multiplied by a geometric factor accounting for the solid angle subtended by each window to find the total number of expected photons, which is then scaled by the transparency of the meshes and the transmission efficiency of MgF$_2$. 
This expectation is then Poisson fluctuated to obtain a predicted light-yield spectrum. 

\begin{figure}[t]
\centering
\includegraphics[width=0.95\linewidth]{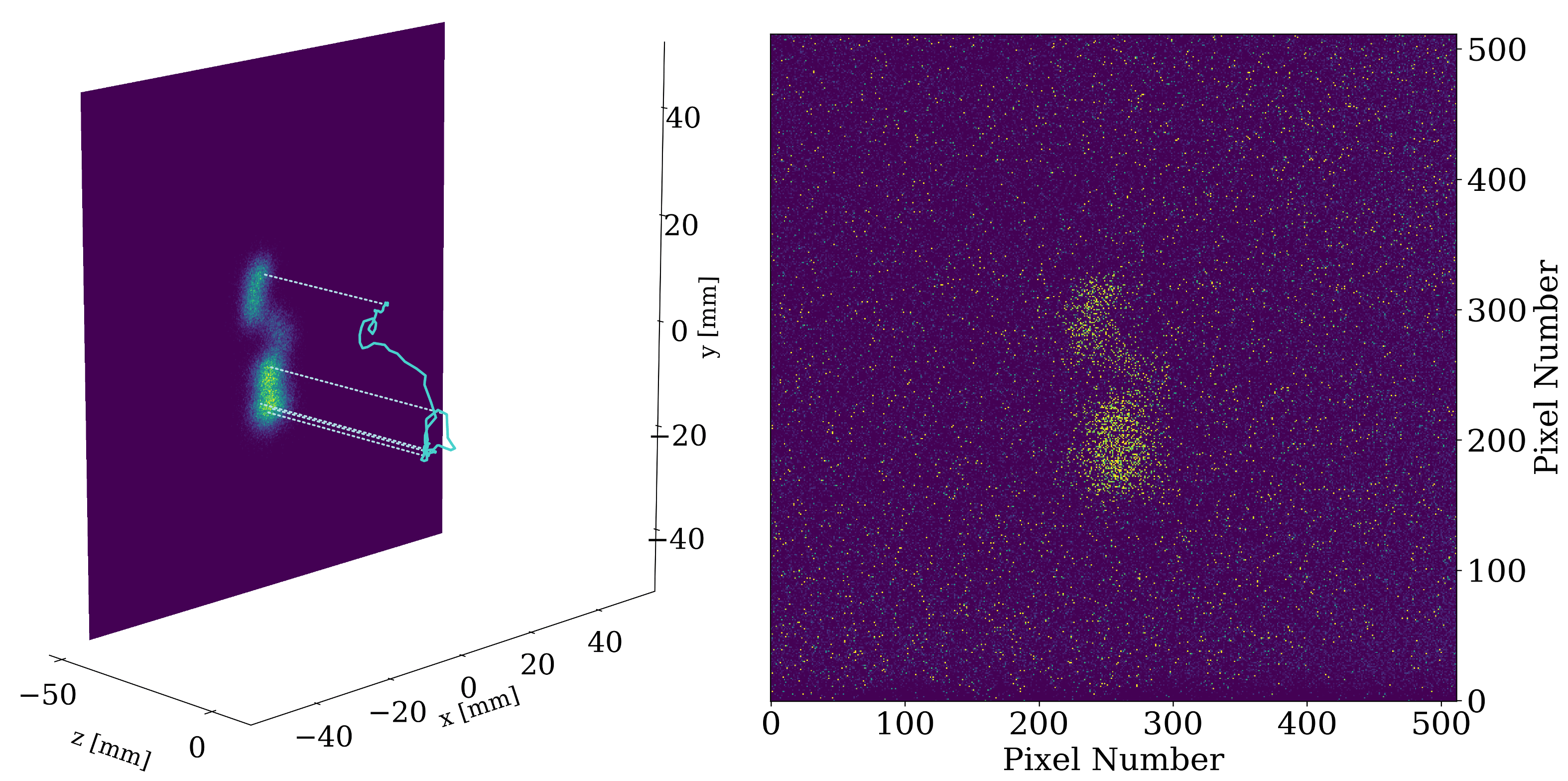}
\caption{One simulated camera captured event track, consisting of two beta decays. 
Left: Simulated 2D electron distribution. 
Right: A simulated camera image of the same event, generated via the techniques in Section \ref{sec:image_sim}.}
\label{fig:3d_truth+sim_image}
\end{figure}

\subsection{Optical track formation and camera simulation}
\label{sec:image_sim}
After predicting the total light yield per event at each window, track images are simulated as they would appear on the EMCCD camera. 
This requires simulation of both the pixelized image after propagation through detector optics as well as the simulation of the random CCD noise background. 

To predict the event sizes and shapes on the camera CCD, geometric optics propagate the simulated event through the optical path including both VUV and visible lenses, projecting the final image on the camera image plane.  
Each simulated frame contains a Poisson-fluctuated number of alphas and betas. 
The mean of this Poisson distribution was calculated using the event rates reported in Section~\ref{sec:calibration} to be 0.7 events per frame.

Effects due to the effective resolution of the imaging system and the gain of the image intensifier are incorporated. 
For the image intensifier, the effective resolution is 9.26 $\mu$m$^{2}$ or the equivalent of 215983 pixels per cm$^2$, sufficiently fine that it has no  noticeable impact on image blurring.  
Effects of spherical aberration and Bokeh were also studied and found to be negligible. 
After applying the gain and quantum efficiency of the image intensifier, the image output from the phosphor screen is then scaled by the solid angle to the transfer lens, and re-binned to match the EMCCD resolution of 512 pixels$^2$, with each pixel-bin value weighted by 0.9 to account for the camera quantum efficiency. 
The result is a histogram of photons per pixel on the CCD. 
Using the value of each bin as the mean expected number of photons detected, these values are sampled from a Poisson distribution to simulate a random integer number of photons at each pixel.
This process is illustrated in the left panel of Fig.~\ref{fig:3d_truth+sim_image}, showing the simulated particle trajectory and how that maps to a two-dimensional pixelated image. 
\begin{figure}[t]
\centering
\includegraphics[width=0.99\linewidth]{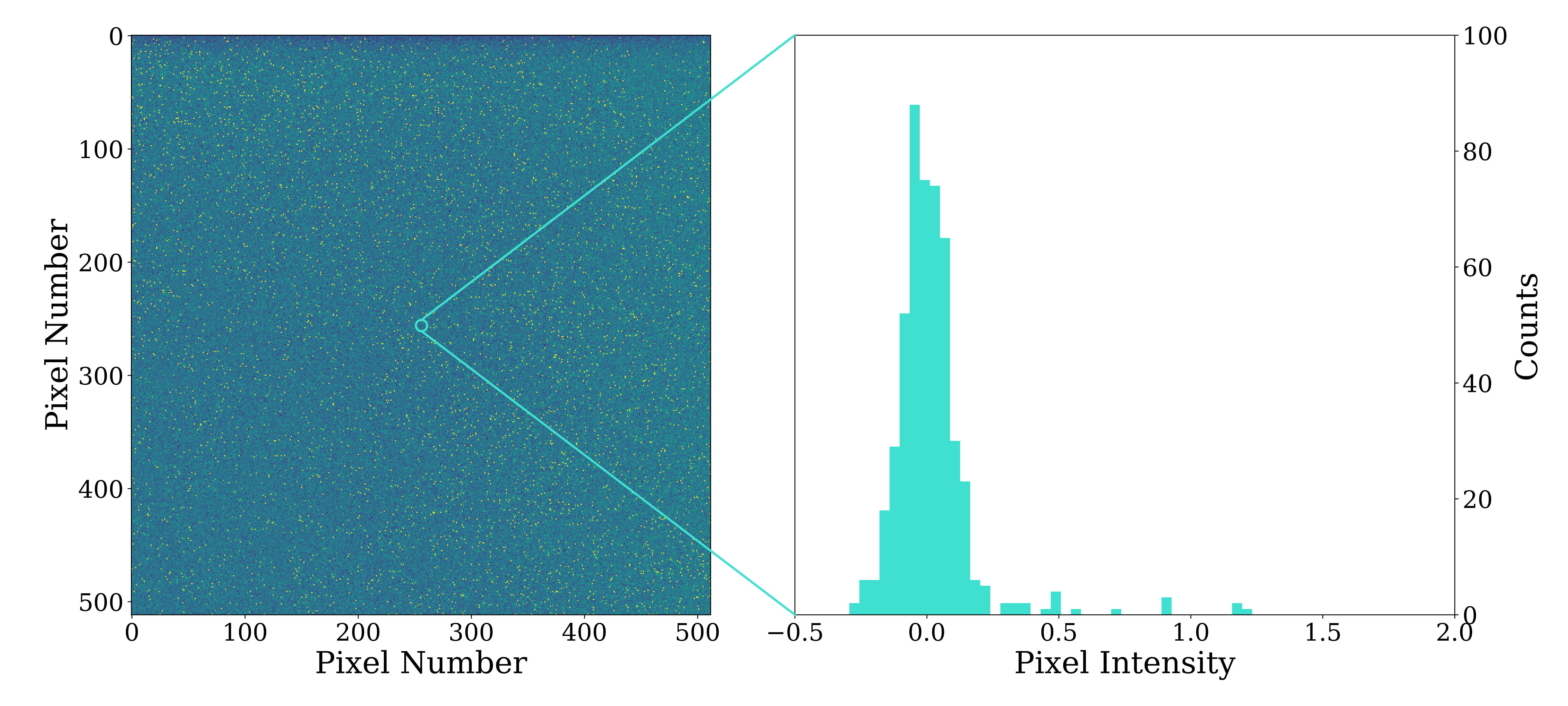}
\caption{An example data-driven CCD noise background.
The histogram inset shows the distribution that the pixel value was randomly selected from in this location.}
\label{fig:Sim_noise_distro}
\end{figure}

\begin{figure}[t]
\centering
\includegraphics[width=0.99\linewidth]{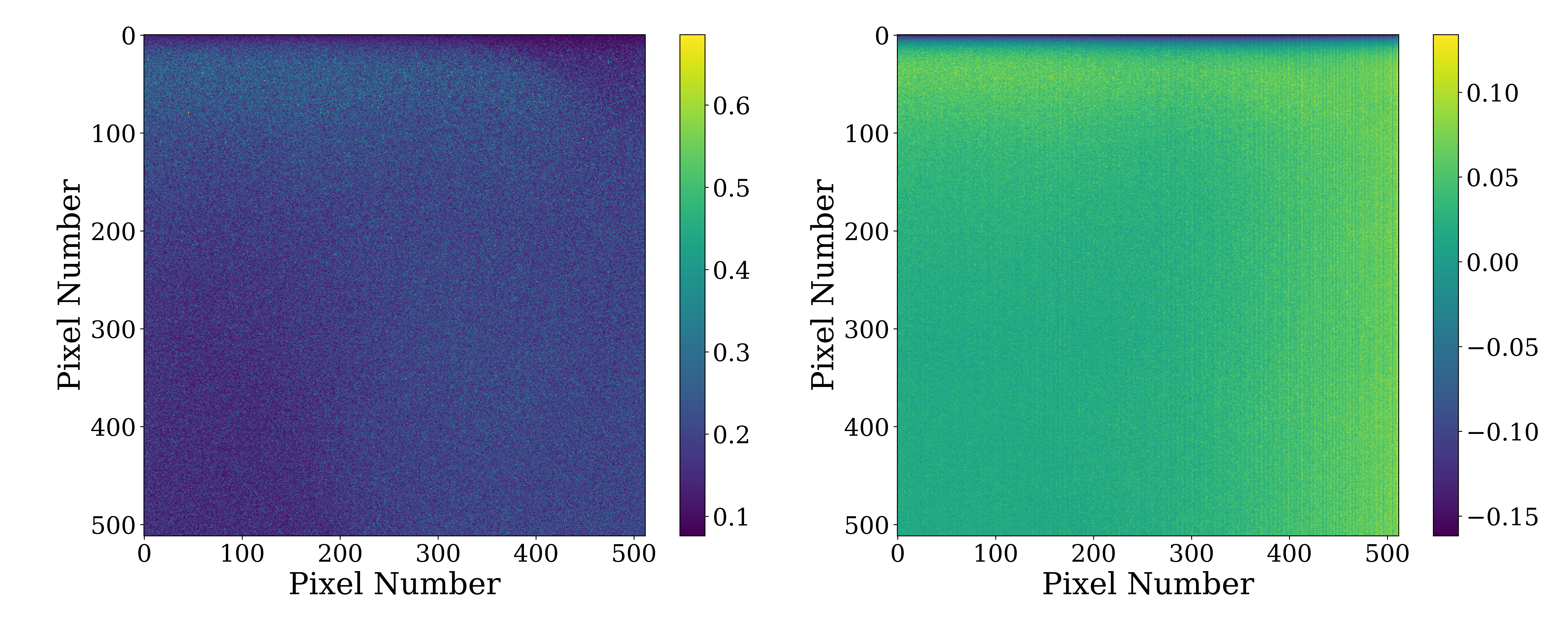}
\caption{Standard deviation (Left) and mean (Right) of the dark image sequence for each pixel.}
\label{fig:Sim_noise_sigma}
\end{figure}
The final step is the addition of simulated CCD noise. 
In order to recreate a realistic background spectrum, a data driven method is used based on pixel noise from a sequence of 500 dark images from the same camera.
From this stack of 512$\times$512 element pictures, a pixel-wise histogram of the noise charge distribution is made.  
The noise in each pixel of the simulated images is randomly sampled from these distributions, as shown in Fig.~\ref{fig:Sim_noise_distro}, with the corresponding standard deviation and mean of each pixel location shown in Fig.~\ref{fig:Sim_noise_sigma}.  
Row-wise and column-wise correlations between noise charge were measured and found to be present at the few-percent level.  
These effects are neglected in the current simulations.
The effects of reflections on image quality were studied in a full optical photon tracking simulation and found to add to the image mostly uniform light across the pixels at the 10~\% level. 
We ignore the reflections in the presented simulated images.
A fully simulated image, including the added noise profile, is included in the right panel of Fig.~\ref{fig:3d_truth+sim_image}.

\section{Experimental Results\label{sec:ExperimentalResults}}

This section presents the results of three experimental studies of various parts of the system. 
First, an {\em ex-situ} test of the imaging optics used to study the positioning of elements and the sensitivity of focus to the various positions; second, a study of the light yield of the TPC system without imaging optics using PMT-based light collection; and third, results from operating the complete system of imaging optics coupled to the time projection chamber.

\subsection{External optical tests\label{sec:OpticsTests}}

\begin{figure}[!t]
    \centering
    \includegraphics[width=0.99\linewidth]{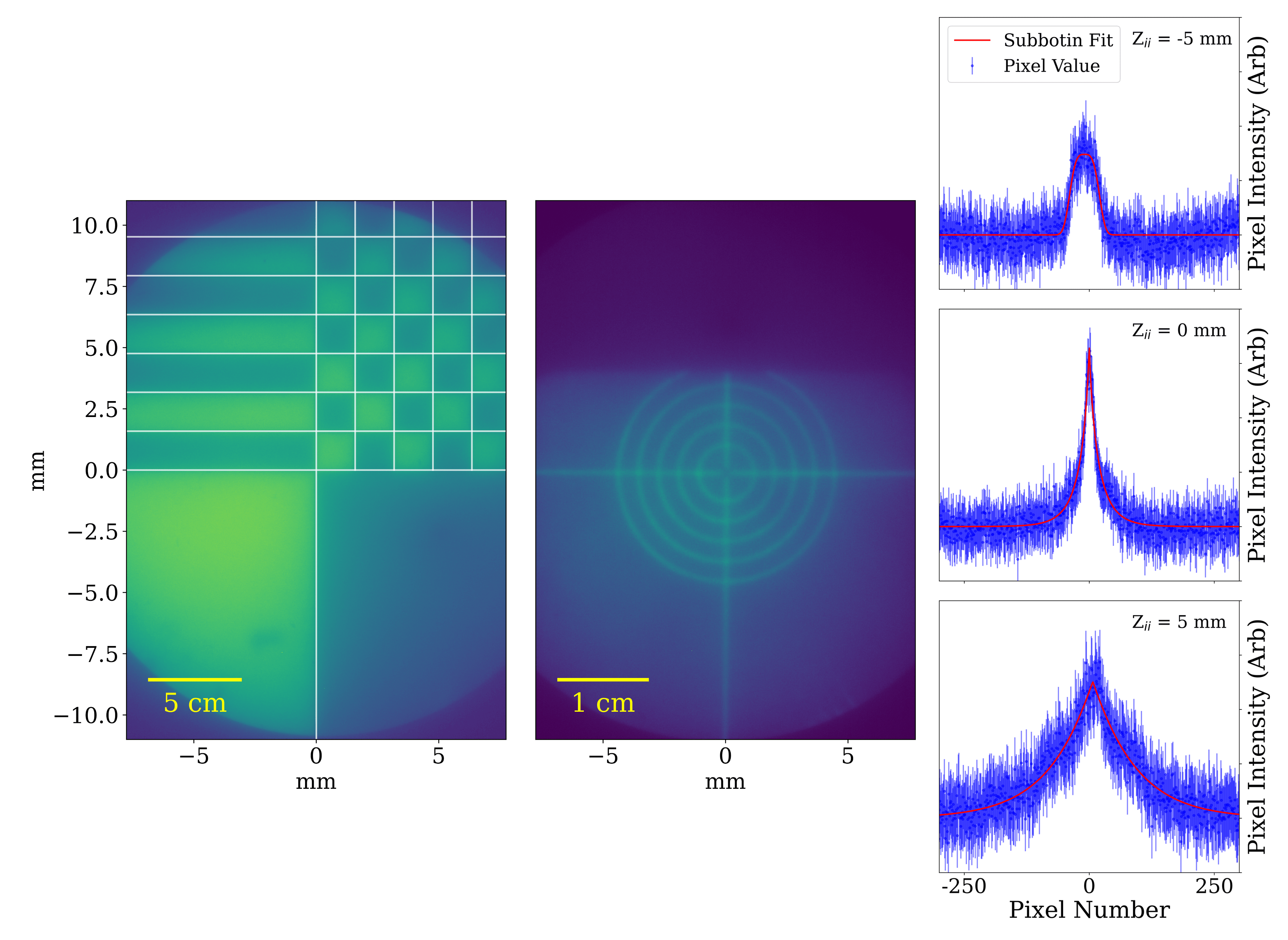}
    \caption{Left and center: Image of the ex-situ targets as imaged by the FLIR Firefly camera and the image intensifier. 
    Right: The averaged profile of the target line with the image intensifier at varying on-axis locations. 
    The red line shows the function fit to these values using Eq.~\ref{equation:Subbotin}. 
    This process was repeated at each recorded position in Fig.~\ref{fig:Bullseye_focus} for both moving optical components.  }
    \label{fig:Target_imgs}
\end{figure}

Prior to installation, the full optical system was prototyped {\em ex-situ} to verify the optical geometry, test the sensitivity of image resolution to various focal distances, and investigate the scaling behaviour of the optical system as it is applied to larger or smaller EL regions.   
These tests aim to quantify the resolution of the image formation optics in a way that is not limited by the pixelization of the EMCCD camera, so a higher resolution FLIR Firefly 1.6 Mpixel CMOS camera was used in place of the Hamamatsu EMCCD. 
These tests were performed using a low-intensity 420~nm blue LED, for which the image intensifier QE is 22.5~\%.

Two targets were used to quantify image resolution achievable with the image intensifier and associated optics and to constrain spherical aberrations, field curvature and other image distortions.  
The first was a square, printed 30~cm wide test image with four quadrants: one white, one black, one with a checker-board pattern, and one with lines alternating back and white.
This target size was chosen to reflect the EL-region diameter of the larger-scale NEXT-CRAB detector, where in principle larger spherical distortions are expected, presently under construction at Argonne National Laboratory.  
After focusing the optics onto this target, images of the target were collected, with an example shown in the left panel of Fig.~\ref{fig:Target_imgs}, and overlaid with a series of vertical and horizontal lines to check that the image is not distorted at this scale. 
No spherical aberrations that would be resolvable within the pixelization of the EMCCD camera were observed.

\begin{figure}[t]
    \centering
    \includegraphics[width=0.99\linewidth]{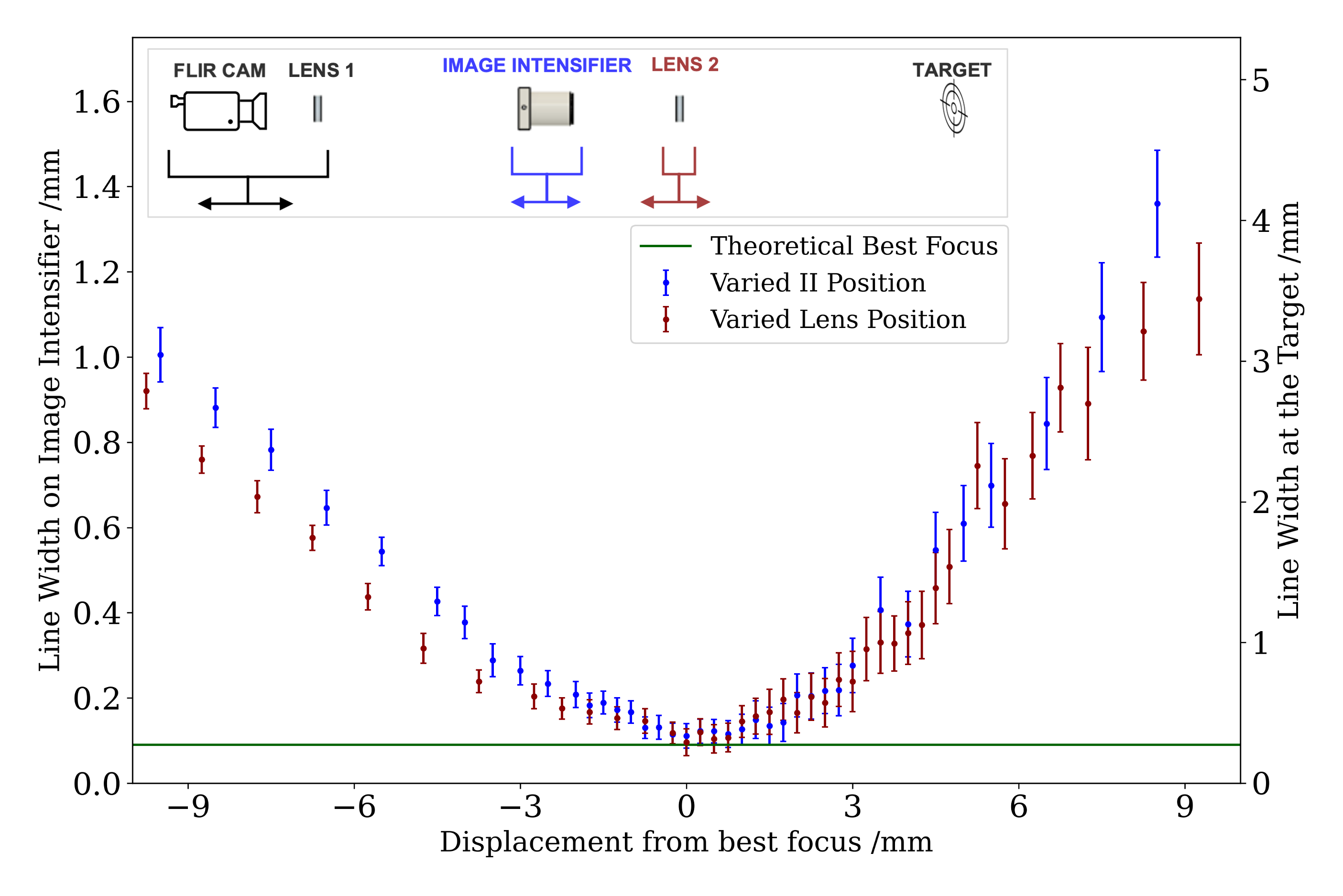}
    \caption{The $\sigma$ of the Subbotin distribution (Eq.~\ref{equation:Subbotin}) fit to the line on the LCPA1 plate plotted as a function of the optical component's position. 
    The left y-axis shows the line width on the image intensifier, while the right y-axis shows the width at the target scale. 
    The green line shows the actual line width, 0.25~mm, and how wide it would be on the image intensifier screen assuming no resolution losses, while blue (red) points show the trend for the image intensifier (Lens) position. 
    The error-bars here are the fit uncertainty of $\sigma$ to the Subbotin. 
    Inset is the diagram showing the movable parts of the optical system.
    }
    \label{fig:Bullseye_focus}
\end{figure}

To quantify the achievable image resolution and hence study the sensitivity to focal positions of various optical elements, a second target was used, a Thorlabs LCPA1 cage alignment plate with 0.25~mm white lines.  
The system was brought into focus using sliding optical stages on micrometer screws coupled to each optical element. 
A focused image of the LCPA1 plate is shown in the central panel of Fig.~\ref{fig:Target_imgs}.  
The width of these lines on the image plane is measured as a proxy for image resolution.  
Each image collected was analyzed by extracting rows of pixels perpendicular to the vertical line in the right center of Fig.~\ref{fig:Target_imgs} at seven distinct positions.
These pixel intensities are averaged to assess the image point spread function. 
To measure the quality of focus at each position, the data were fit with the following empirically derived function,
\begin{equation}
\label{equation:Subbotin}
    I(x) = ax^3 + bx^2 + cx + d + \frac{\alpha\eta}{2\sigma\Gamma\left(\frac{1}{\eta}\right)}\exp\left[-0.5\left(\frac{|x-\mu|}{\sigma}\right)^\eta\right].
\end{equation}
In this expression, parameters $x$ and $I$ represent pixel coordinate and intensity respectively; $a,b,c,d$ are fit parameters that define a smoothly varying function that well models the diffuse light background, $\alpha,\mu,\eta,\sigma$ are fit parameters of the Subbiton distribution providing an acceptable fit to the line shape at each tested focal distances.  
Eq.~\ref{equation:Subbotin} was fit at various positions of the lens and image intensifier to provide a measure of the line width in each optical configuration. Three fit examples are shown in Fig.~\ref{fig:Target_imgs}, right.  
The width parameter $\sigma$ is plotted as a function of the position of each optical element in Fig.~\ref{fig:Bullseye_focus}.  
At the best focus position for both optical components, the $\sigma$ of this system is 0.10~mm at the image intensifier output, 0.004\% of the output area, corresponding to an X-Y resolution at the target distance of 0.27~mm.  
There is similar sensitivity to the displacement of either the image intensifier or the imaging lens. 
A positioning precision of around 1~mm is found to be suitable for achieving optimal image quality in the {\tt NEXT-CRAB-0} system.

\subsection{Light yield characterization}
\label{sec:light-yield}

\begin{figure}[t]
\centering
\includegraphics[width=0.99\linewidth]{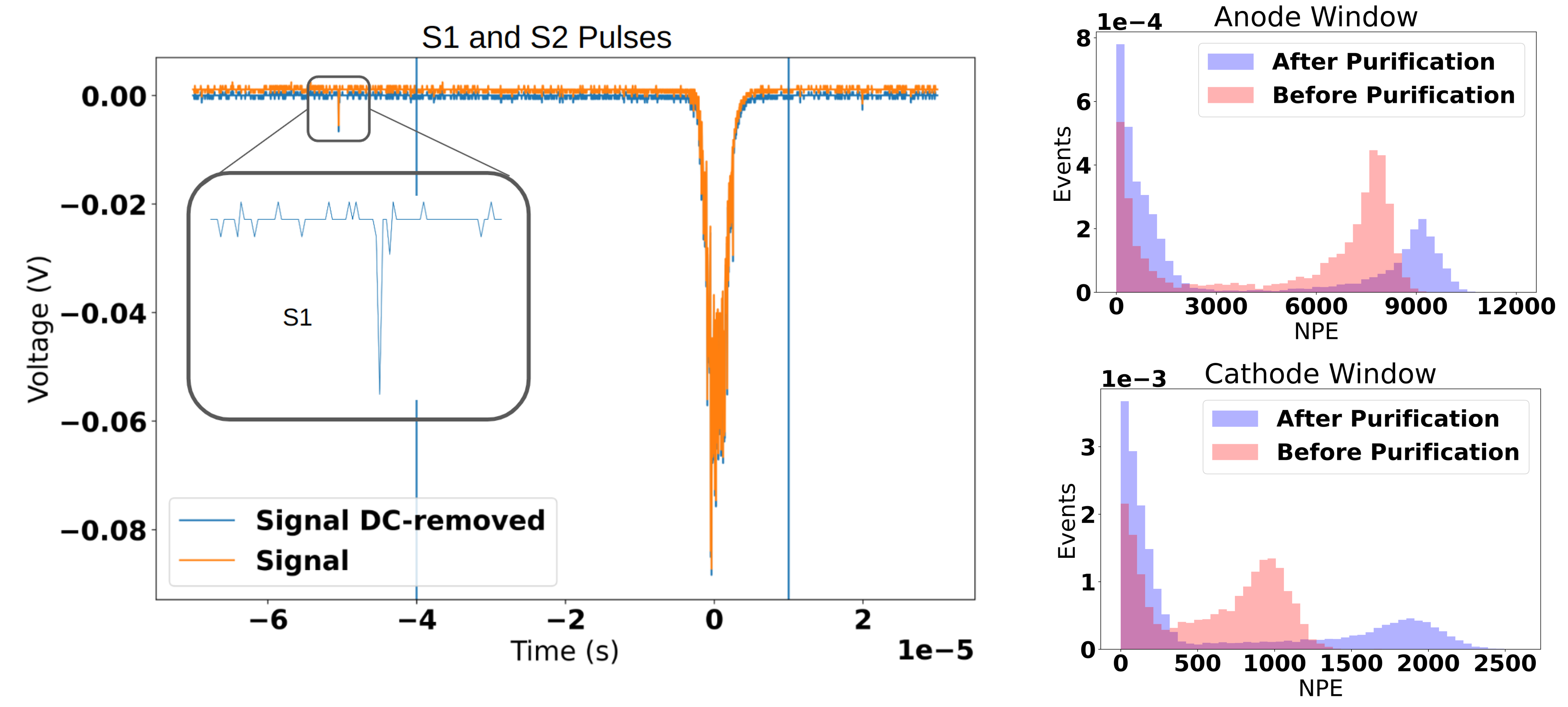}

\caption{Left: Example of a single triggered data event showing S1 and S2 pulses. 
The raw signal (orange) is corrected by a DC offset factor to get the DC-removed signal (blue). 
The two blue lines represent the integration window of the S2 signal. 
Right: total detected light yield distributions at each PMT, before and after gas purification.}
\label{fig:SinglePulseAnalysis}
\end{figure}

In the second set of studies, the light yield at each window was characterized with the detector filled at 9.7~bar xenon gas. 
For this study, the image intensifier is replaced with a second Hamamatsu R7378A PMT installed in the image intensifier housing at the cathode end of the detector. 
This PMT is labelled as ``Cathode Window.''  The original PMT at the anode end is also operated and referred to here as ``Anode Window.'' 
The distance from Cathode Window to the EL region midplane is measured to be 35.5$\pm{0.5}$~cm , and from Anode Window to EL region midplane is 13.5$\pm{0.5}$~cm. 
Both PMTs are powered by a pair of Hamamatsu C9525 high voltage power supplies.
PMT bias voltages were brought slowly to 1200 V and left to stabilize for at least 1 hour prior to data taking.  
The single photoelectron response of each PMT was characterized {\tt in-situ} with a pulsed LED, as described in Appendix B.

The signal waveforms were collected for both PMTs using a three stage trigger on the oscilloscope. 
The three stage trigger requires coincidence between an Anode Window pulse with both S1 and S2 signals with a Cathode Window S2 signal.  
This way, waveforms are obtained only for events with a significant number of photons detected on both PMTs at the same time and clear S1 and S2 pulses. 
A typical triggered event showing clear S1 and S2 signals in both PMTs is shown in Fig.~\ref{fig:SinglePulseAnalysis}, left.

Data were recorded both after an initial fill through the cold getter and then again after 24 hours of purification (Fig.~\ref{fig:SinglePulseAnalysis}, right).  
All runs show a clear alpha peak and a beta decay spectrum at lower charge, with the high purity runs found to exhibit significantly increased light yields.  
The light yield was found to stabilize without noticeable improvement after this initial 24~hr of circulation.

To quantify the number of S2 photons and hence benchmark the light yield, the DC offset was removed from the recorded scintillation pulses and then a 30~$\mu$s region around the trigger was integrated (blue lines on Fig.~\ref{fig:SinglePulseAnalysis}, left) to obtain the area of each S2 event. 
This area is divided by the measured PMT gains to obtain the photoelectron distribution to be compared against expectations from simulations. 

 \begin{figure}[!ht]
\centering
    \includegraphics[width=0.80\linewidth]{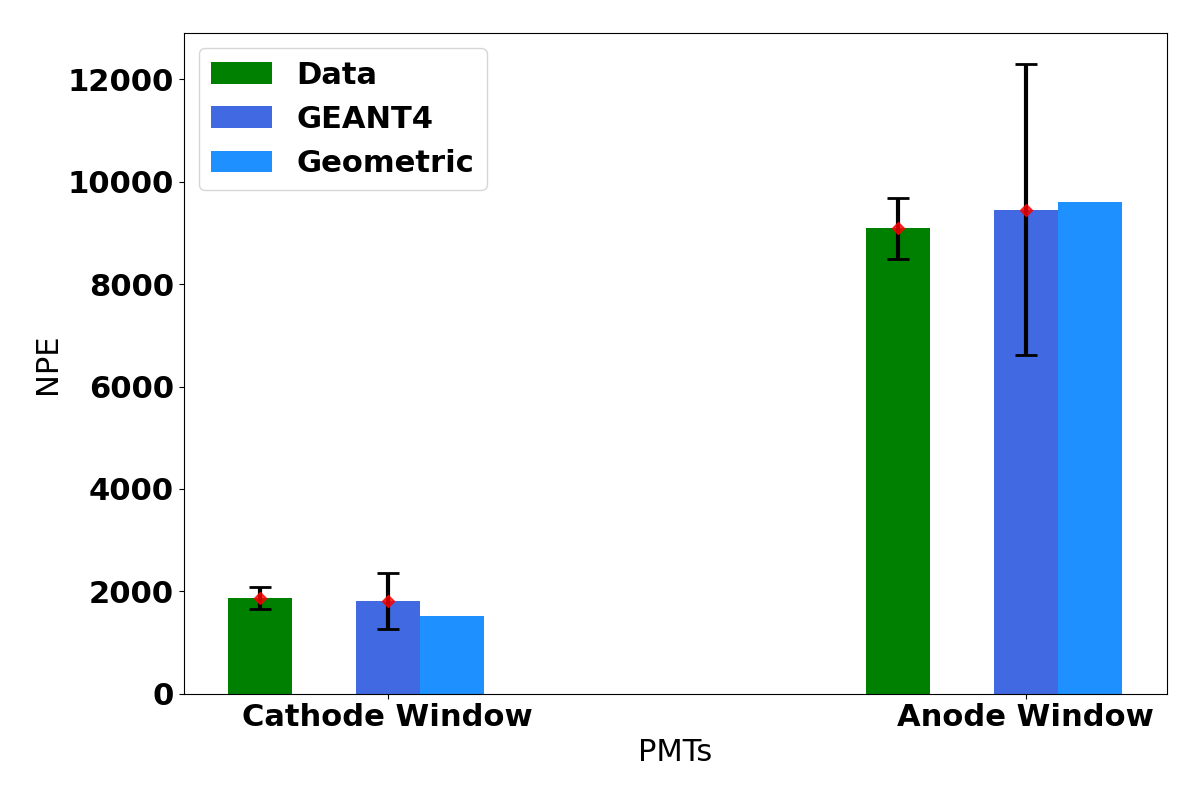}
\caption{Alpha light yield comparison between the simulation and data from the high purity run. }
\label{fig:S2_Comparison}
\end{figure}

The mean S2 light yields due to 5.3 MeV alpha particles are compared between MC for Anode and Cathode Windows in  Fig.~\ref{fig:S2_Comparison}, obtained by fitting a Gaussian curve to the observed S2 charge distribution.  
The error bar on the data reflects the width of this fitted Gaussian, while the one on the {\tt GEANT4} simulation is the uncertainty associated with the $\pm 30\%$ radiative cathode sensitivity uncertainty of the PMTs~\cite{PMT}.
The prediction for the geometric optics calculation is provided for comparison to the {\tt GEANT4} simulation, pinned to the same quantum efficiency, so does not carry an independent error bar.  
These results are presented in the units of NPE/MeV, where NPE is the number of photoelectrons, in Table~\ref{table:S2_LightYields}. 

\begin{table}[h!]
\begin{center}
\centering
\begin{tabular}{ |p{2cm}||p{5cm}||p{5cm}|  }
 \hline
 Method & Anode Window $(\mathrm{NPE}/\mathrm{MeV})$& Cathode Window $(\mathrm{NPE}/\mathrm{MeV})$ \\
 \hline
 Data &  $1.72\times10^{3} \pm {0.11\times10^{3}}$ & $3.52 \times10^{2} \pm{0.40\times10^{2}}$ \\
 GEANT4 & $1.78\times10^{3} \pm {0.54\times10^{3}}$ & $3.41 \times10^{2} \pm{1.0\times10^{2}}$ \\
 Geometric & $1.81\times10^{3} $ & $2.88  \times10^{2} $ \\
 \hline
\end{tabular}
\end{center}
\caption{Measured and estimated alpha S2 light yields.} \label{table:S2_LightYields}
\end{table}

\subsection{Camera-based tracking}
\label{sec:signal-proc}

The camera pixelization is fine relative to the expected width of the EL light arriving at the camera after signal broadening through diffusion, which means the detected events are sparse in terms of photons per pixel. 
To enhance the visibility of signal activity and suppress noise, a 2D Wiener filter~\cite{WienerFilter} is employed, described in detail in  Appendix C.  

Data were taken using the $^{210}$Pb calibration source and images were recorded and passed through the Wiener filter.  
The images were captured with a 28 ms camera exposure and full EM gain factor of 1200$\times$.  
The image intensifier was operated at its maximal gain. 
Data were collected for a varying electric field intensity within the EL region and visually the best track signals were obtained with a reduced electric field of 1.43~kVcm$^{-1}$bar$^{-1}$.  
Attempts to exceed this field strength resulted in sparking to the vessel walls that set an upper threshold on the EL gain, a roughly 16\% reduction in field strength from that of NEXT-White's Run II \cite{Monrabal:2018xlr}. 
The CRAB-0 prototype aims to produce clear particle tracks, and so sufficient high voltage must be placed over the EL region to cross the EL light threshold of 0.73~kVcm$^{-1}$bar$^{-1}$.  
The working fields in the EL region are not as large as in full-scale demonstrator detectors that employ xenon gas, however, since the distance to the vessel walls is shorter in this geometry, resulting in the breakdown between the EL and the vessel wall at 1.43~kVcm$^{-1}$bar$^{-1}$.

In these conditions, our simulations show that the yield of camera photoelectrons per predicted primary ionization electron is expected to be around 0.070.  
This number accounts for the EL gain (648~ph/e), mesh transparencies (0.805), first lens solid angle (1.33$\times$10$^{-4}$~sr), MgF$_2$ transmission efficiency (0.87$\pm0.02$),  image intensifier quantum efficiency (0.22) and gain (3000), efficiency of transfer optics to the camera (1.95$\times$10$^{-3}$), and camera quantum efficiency (0.9).  
A systematic uncertainty is not provided on the image intensifier gain or quantum efficiency, but we estimate that they are likely at least 10\% each, which add in quadrature to the 2\% uncertainty on the MgF$_2$ transparency and a 5\% total uncertainty from geometric effects to give a central value and total error budget of $0.070\pm0.011$\%. 

\begin{figure}[t]
\centering
\includegraphics[width=0.8\linewidth]{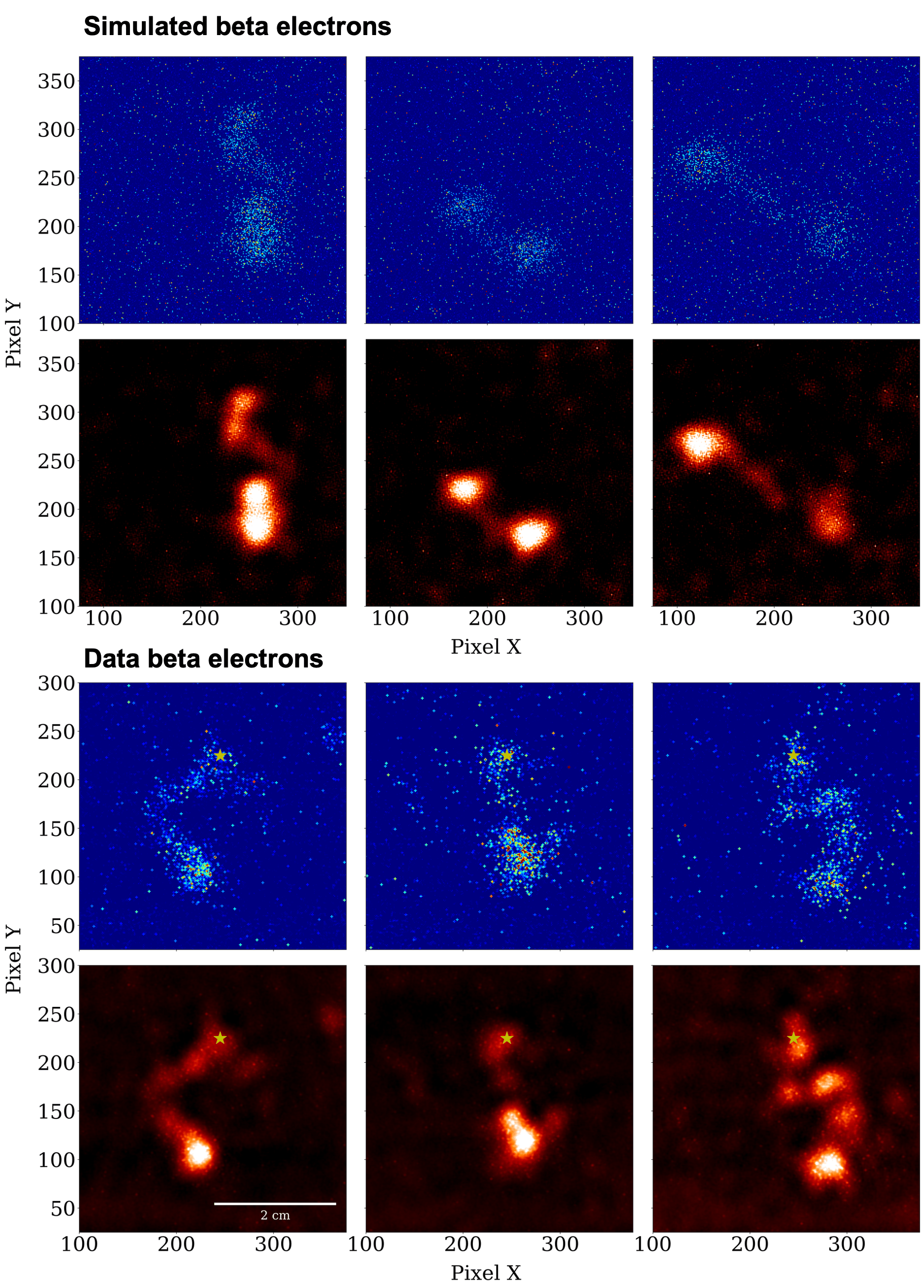}
\caption{Simulated (Top) and detected (Bottom) beta electrons in {\tt NEXT-CRAB-0}. 
The blue images are before filtering and the red images are after the filtering protocol is applied. }
\label{fig:sim_betas}
\end{figure}

\begin{figure}[t]
\centering
\includegraphics[width=0.8\linewidth]{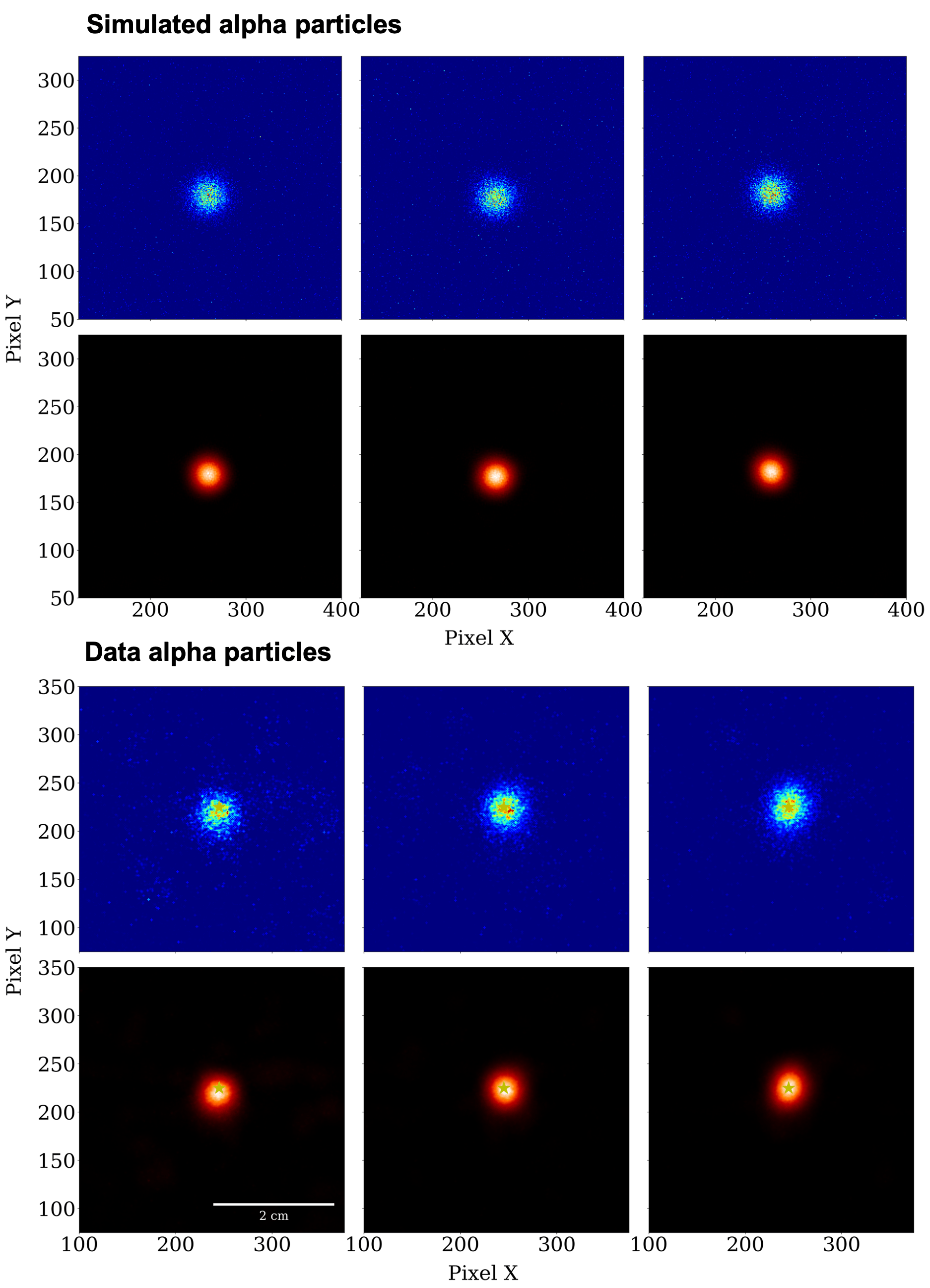}
\caption{Simulated (Top) and detected (Bottom) alphas in {\tt NEXT-CRAB-0}.  
The blue images are before filtering and the red images are after the filtering protocol is applied. }
\label{fig:sim_alphas}
\end{figure}

The expected number of alpha particles per frame has been estimated in two ways. 
First using the alpha rate as observed from the PMT and the exposure time of 28~ms, we obtain $\lambda_{\alpha,1}=0.35\pm0.01$. 
Second, using the fraction of alpha-free camera images in the signal data set we obtain $\lambda_{\alpha,2}=0.41\pm0.01$ (statistical uncertainties only). 
An image was determined to be alpha-free by taking the integrated pixel intensity in a $80\times80$ pixel square around the needle source, and applying an intensity cut to exclude any images containing an alpha within that window. 
We take the geometric mean of these two measurements as our estimate, $\lambda_\alpha=0.38\pm0.02$.   
Using this number and taking the background-subtracted integration of the average charge in the $\alpha$ region of the images, the measured yield of camera photoelectrons per predicted primary ionization electron in data is found to be 0.064$\pm0.003$.  
Agreement of this metric between simulation and data suggests that the imaging system performance is well modeled and understood.

\begin{figure}[t]
\centering
\includegraphics[width=0.8\linewidth]{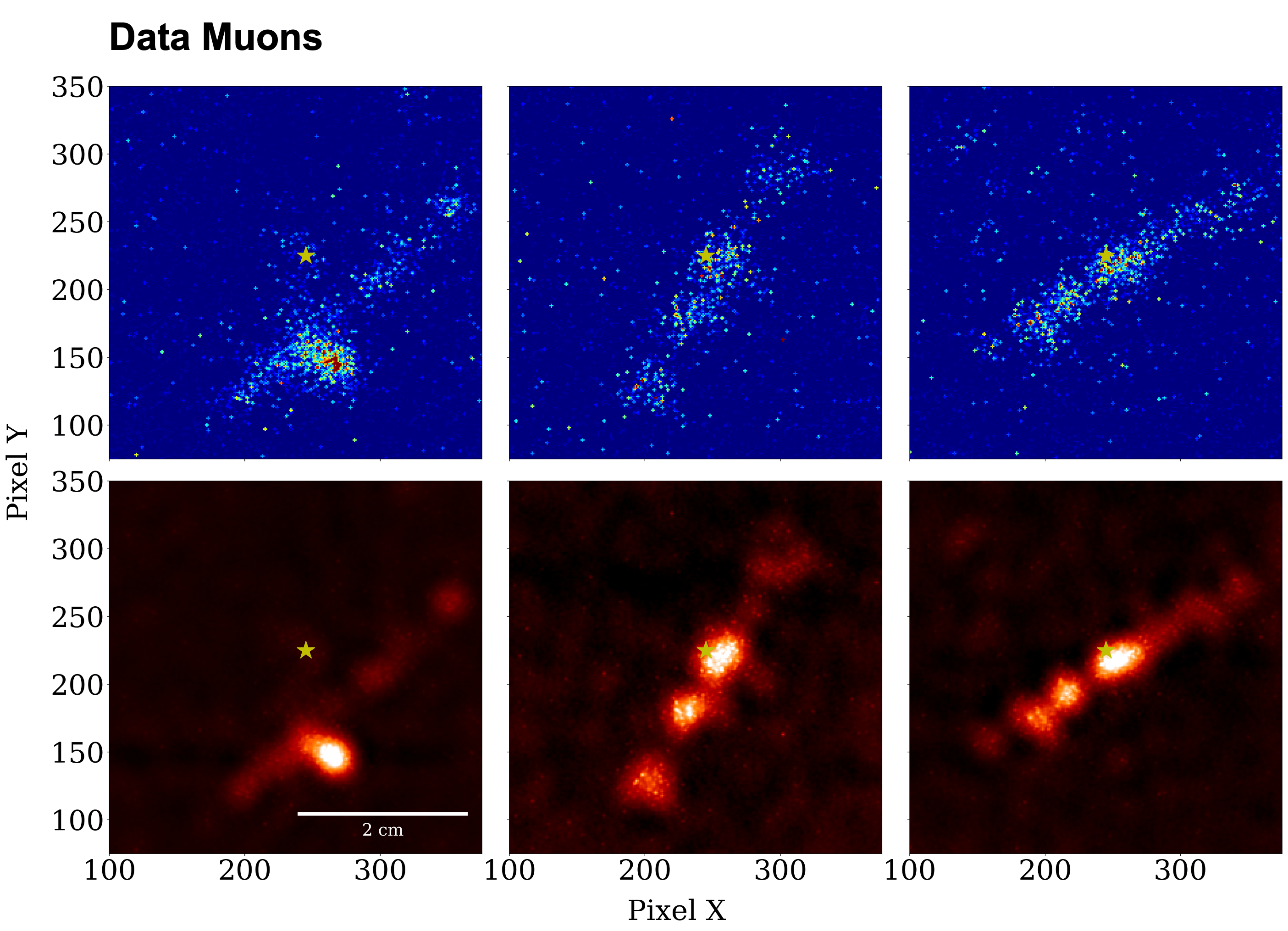}
\caption{Three candidate muon events captured during data collection. 
The blue images are before filtering and the red images are after the filtering protocol is applied. }
\label{fig:muons}
\end{figure}

Fig.~\ref{fig:sim_betas} compares beta electron simulations (top) with detected candidates~(bottom). 
Similarly, Fig.~\ref{fig:sim_alphas} shows candidate single alpha data events compared against simulated images.  
Qualitatively these images appear remarkably similar, and a full topological analysis using NEXT-CRAB data will be the subject of a future publication.    
Quantitatively, the observed photon yields also agree well with first-principles simulations. 
These events represent the ``first light'' from the  {\tt NEXT-CRAB-0} demonstrator, providing the first VUV image intensified tracking of EL amplified MeV-scale beta tracks.   
We also observe a number of cosmic ray muon events in the dataset, some examples (likely overlaid on alpha or beta tracks) are shown in Fig.~\ref{fig:muons}.

\section{Conclusion}
\label{sec:Conclusion}

The search for $0\nu \beta \beta$ decay presents many challenges, chief among them the separation of signal events from pernicious backgrounds due to cosmogenic activity and material radioactivity. 
Given what is presently known about neutrino masses and mixing,  observation of this decay will  require a ton- or multi-ton-scale experiment with excellent background rejection (<1 count/ROI/ton/yr) if the neutrino mass ordering is inverted.  
If the neutrino masses are normally ordered, even more impressive background rejection levels will be required.  
The NEXT collaboration aims to meet these challenges using scalable, ultra-low background xenon gas time projection chambers with EL amplification. 

The NEXT-CRAB technique is an emerging technology to realize topological tracking while minimizing system complexity, radioactivity and cost at large detector scales.   
The removal of SIPMs from the interior of the detector removes a significant radiogenic background contributor, and through the addition of mirrors to the optical path, cameras need not align directly with the vessel view port and shielding can be added to remove the camera as a source of radioactivity. By reducing the real estate occupied by the tracking system to one or a few MgF$_2$ lenses situated behind a low-voltage cathode, the technique also facilitates the inclusion of future barium tagging systems that are expected to occupy most of the detector cathode in future experiments. 

This paper has presented the ``first light'' from the {\tt NEXT-CRAB-0} demonstrator, providing clear images of alpha, beta and muon events across a 7.2~cm diameter EL  time projection chamber operating with 9.7~bar xenon gas.  
A detailed {\tt GEANT4} Monte Carlo simulation of the system has been developed, and augmented with both geometric optics and photon-tracking methods to model light propagation through the imaging system through to image formation. 
Light yields at photomultiplier tubes at both detector windows were found to be consistent with expectations from simulations within systematic errors that are dominated by the unknown quantum efficiency of the PMT photocathode at 172~nm.  
First-principles predictions of light yields at the camera predict an expected photoelectron yield per primary electron of 0.070$\pm$0.011, which compares well with a measurement made using $^{210}$Po alpha particles yielding 0.064$\pm$0.003.

The experiment presented in this paper had important limitations, most notably the small fiducial volume prohibiting containment of events near the Q-value for $0\nu\beta\beta$ of $^{136}$Xe, as well as the restriction of tracking only in 2D imposed by the speed of the presently available EMCCD camera. 
The coming phase of NEXT-CRAB that is under construction at Argonne National Laboratory will address both of these issues. 
The larger scale CRAB system has a 30~cm diameter EL region and holds 22~kg of xenon at 15~bar.  
A TimePix-based camera coupled to a double MCP image intensifier will provide a gain that is comparable with that achieved in {\tt NEXT-CRAB-0}, and improve the time resolution of the readout to the point that full 3D event tracking is achievable. 
With the rich track imaging information acquired from such a device,  the performance of topological analyses will be evaluated for the potential to separate single- from two-electron events in the $0\nu\beta\beta$ energy range of interest.  
Should this performance prove to be competitive with the topological identification techniques pioneered by the NEXT collaboration to date, the NEXT-CRAB concept could form the basis for a future ton- to multi-ton scale program searching for $0\nu\beta\beta$ of $^{136}$Xe using ultra-low background xenon gas time projection chambers.

\acknowledgments

This work was supported by the US Department of Energy under awards DE-SC0019054 and  DE-SC0019223, the US National Science Foundation under award number NSF CHE 2004111 and the Robert A Welch Foundation under award number Y-2031-20200401 (University of Texas Arlington).  
FJS was supported by the DOE Nuclear Physics Traineeship Program award DE-SC0022359. 
The NEXT Collaboration also acknowledges support from the following agencies and institutions: the European Research Council (ERC) under Grant Agreement No. 951281-BOLD; the European Union's Framework Programme for Research and Innovation Horizon 2020 (2014–2020) under Grant Agreement No. 957202-HIDDEN; the MCIN/AEI of Spain and ERDF A way of making Europe under grants RTI2018-095979 and PID2021-125475NB , the Severo Ochoa Program grant CEX2018-000867-S and the Ram\'on y Cajal program grant RYC-2015-18820; the Generalitat Valenciana of Spain under grants PROMETEO/2021/087 and CIDEGENT/2019/049; the Department of Education of the Basque Government of Spain under the predoctoral training program non-doctoral research personnel; the Portuguese FCT under project UID/FIS/04559/2020 to fund the activities of LIBPhys-UC; the Israel Science Foundation (ISF) under grant 1223/21; the Pazy Foundation (Israel) under grants 310/22, 315/19 and 465; the US Department of Energy under contracts number DE-AC02-06CH11357 (Argonne National Laboratory), DE-AC02-07CH11359 (Fermi National Accelerator Laboratory), DE-FG02-13ER42020 (Texas A\&M). Finally, we are grateful to the Laboratorio Subterraneo de Canfranc for hosting and supporting the NEXT experiment.

\section*{Appendix A: {\tt LabVIEW} slow controls}

\begin{figure}[t]
\centering
\includegraphics[width=0.8\linewidth]{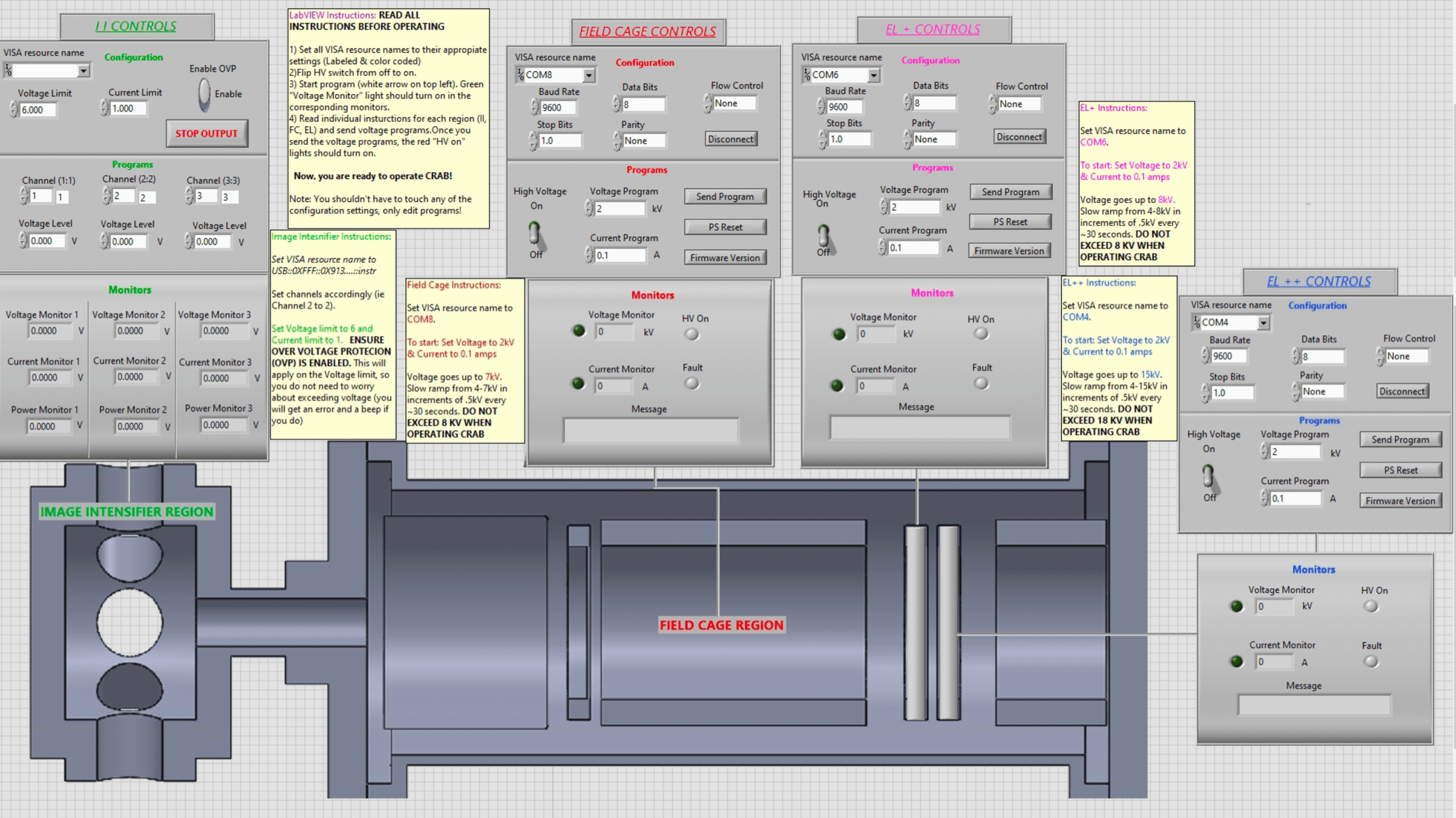}

    \caption{{\tt NEXT-CRAB-0} slow controls front panel. See text for description.}
\label{fig:FrontPanel}
\end{figure}

The three Glassman FJ power supplies that supply detector high voltage, as well as the low voltage controls of the image intensifier, are  controlled and monitored remotely through a bespoke  LabVIEW slow control system.  

The Field Cage and EL+ are both connected to a 12~kV, 10~mA Glassman while the EL++ is connected to a 20~kV, 6~mA Glassman power supply, all connected to a slow controls computer via an Ethernet switch.  
Instances of the three virtual interfaces were modified to incorporate ramping and monitoring functionality and added to the application front panel. 
The image intensifier power and gain settings are controlled by a BK Precision 9132B power supply over USB, which has control blocks embedded into the same front panel. 
The BK Precision's three channels allow for the remote management of the image intensifier's operating voltage, gain control, and ground connection. 
A screenshot of the slow controls front panel is shown in Fig.~\ref{fig:FrontPanel}.

\section*{Appendix B: Single photoelectron calibration protocol}

\begin{figure}[h!]
\centering

\includegraphics[width=0.99\linewidth]{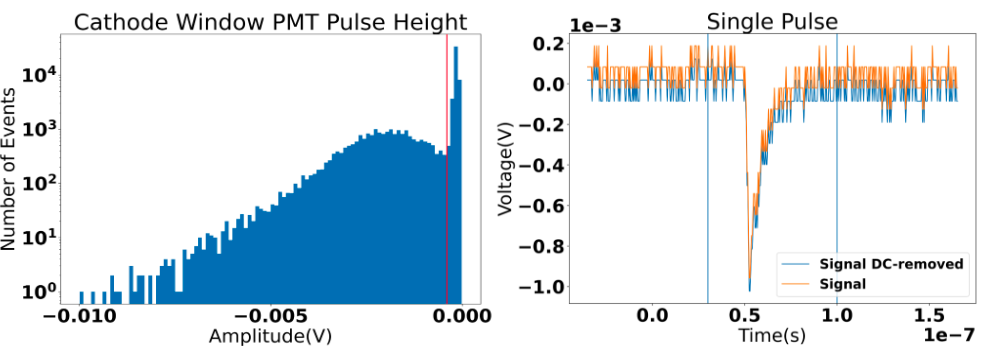}

\caption{Left: Pulse height spectrum for the cathode window PMT. 
The red line indicates a rough cut made to separate the SPE peak from noise. 
Right: Example SPE waveform used in the calibration procedure, before and after baseline subtraction. }
\label{fig:PMTPulseHeight}
\end{figure}

The PMTs used for light yield measurements were calibrated using a 430~nm pulsed LED driven by a function generator (RIGOL 1032Z). 
The LED was coupled to a UV/Visible fiber optic feedthrough with internal fiber placed near the EL region, facing toward the Cathode Window.  
The LED was pulsed at 5~kHz with amplitude 3.5~V and 20~ns pulses.  
This configuration led to a spectrum of primarily zero photoelectron responses on both PMTs, with a small single photoelectron (SPE) population. 
The pulse was used as an external trigger to the oscilloscope and the PMT waveform was recorded. 
An example recorded SPE pulse signal is shown in Fig.~\ref{fig:PMTPulseHeight}, left. 

An example distribution of PMT pulse heights at the cathode window is shown in Fig.~\ref{fig:PMTPulseHeight}, left. 
After making a rough noise cut at -0.4~mV, the DC offset was subtracted from each pulse using pulse pre- and post-samples as a reference. 
The area of the pulse was then integrated between gates which is 70~ns wide shown in Fig~\ref{fig:PMTPulseHeight}, right. 
The distribution of SPE areas is used to determine the PMT gain for each of the two PMTs. 
The Cathode Window PMT gain was estimated to be $ (382\pm 1.83)\times10^{4}$  and the Anode Window PMT gain is $(254\pm 0.987) \times 10^{4}$. 
Based on the observed ratio of 0 to 1 PE pulses, the expected contamination from N$\geq$ 2 PE pulses from Poisson statistics is estimated to be less than 5\%, which would lead to negligible correction to these measurements.

\section*{Appendix C:  Wiener filter image processing}

A Wiener filter is a matrix operation applied in frequency space to the 2D fast Fourier transform of data images.  
The Wiener filter for our application is constructed using a data driven approach as shown in Eq.~\ref{equation:Wiener1},
\begin{equation}
    W_{ij} = \frac{C_{ij} - B_{ij} }{C_{ij}}.
    \label{equation:Wiener1}
\end{equation}
Here, \textit{C$_{ij}$} is the average power spectrum of all the images with observed tracks while \textit{B$_{ij}$} is the normalized average  background power spectrum, with \textit{i} labeling each column and \textit{j} each row in frequency space.

The power spectrum of the observed track and background are related to the discrete Fourier transforms of the images by
\begin{eqnarray}
    C_{ij} &=& \frac{1}{N^{'}} \sum_{a=1}^{N^{'}} \frac{1}{D^2}|FT(I^a)_{i,j}|^2,\\
    B_{ij} &=&f \times \frac{1}{N} \sum_{a=1}^{N} \frac{1}{D^2}|FT(Y^a)_{i,j}|^2,
    \label{equation:Wiener1.2}
\end{eqnarray}  
where \textit{I}  (Y) is an observed signal (background) image with $D^2$ = 512x512 pixels. 
$N$ and $N^{'}$ are the total number of  images that are used. 
$FT(I^a)_{ij}$ and $FT(Y^a)_{ij}$ are the discrete Fourier transforms of the observed images with index ``$a$''.  
$f$ is a normalization factor used to ensure the noise backgrounds on signal and noise runs are of comparable magnitude.

\begin{figure}[t]
\centering
\includegraphics[width=0.99\linewidth]{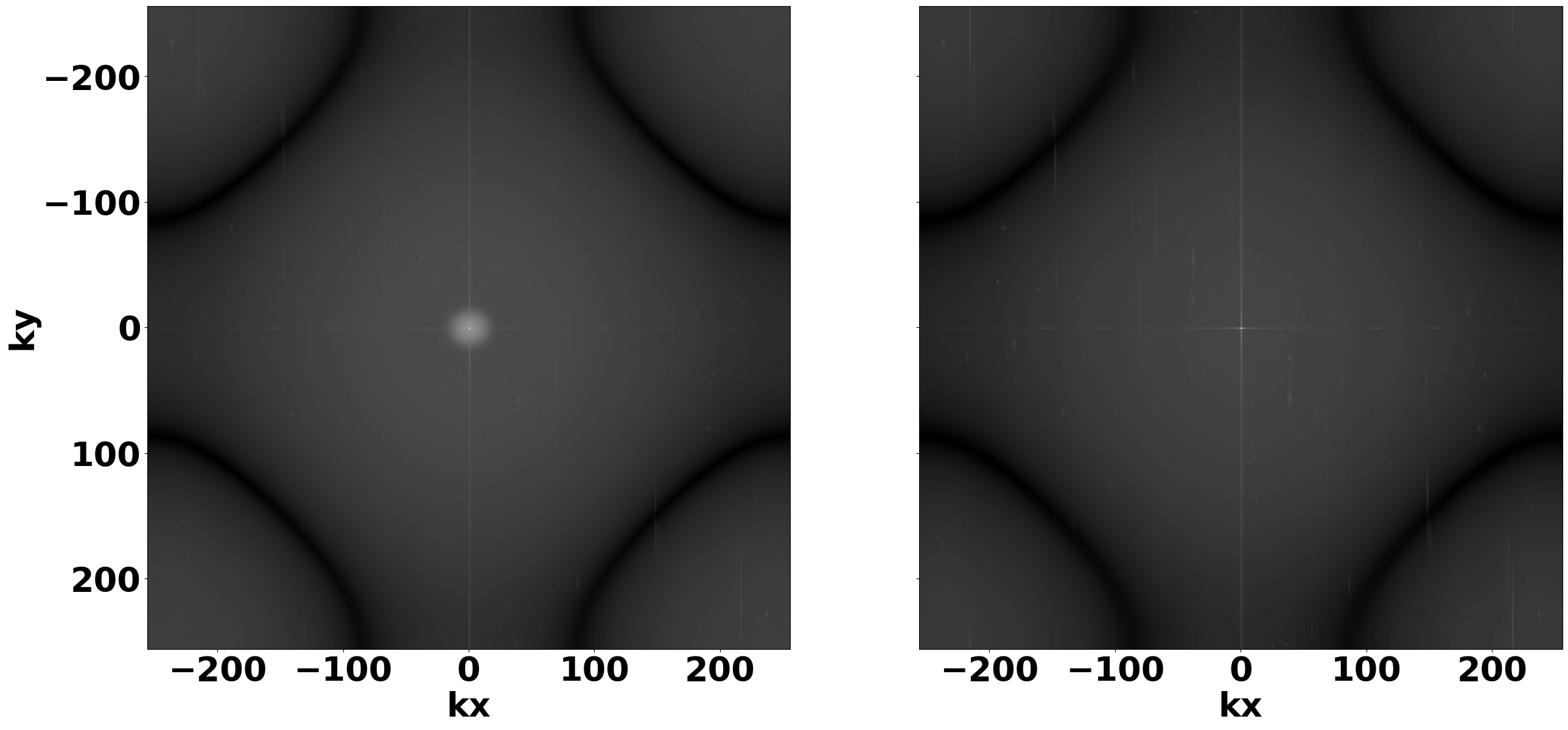}
\caption{Left: 2D average power spectrum of track images.
Right: 2D average power spectrum of normalized background.}
\label{fig:ShiftedAvgImgBg}
\end{figure}

The 2D power spectrum of averaged observed tracks and normalized background are shown in Fig.~\ref{fig:ShiftedAvgImgBg} where~\textit{kx} and~\textit{ky} represent the frequency. 
The high activity seen at low frequencies corresponds to diffused track information, with a diffuse  low intensity background of roughly white noise visible over all frequencies. 

\begin{figure}[t]
\centering
\includegraphics[width=0.99\linewidth]{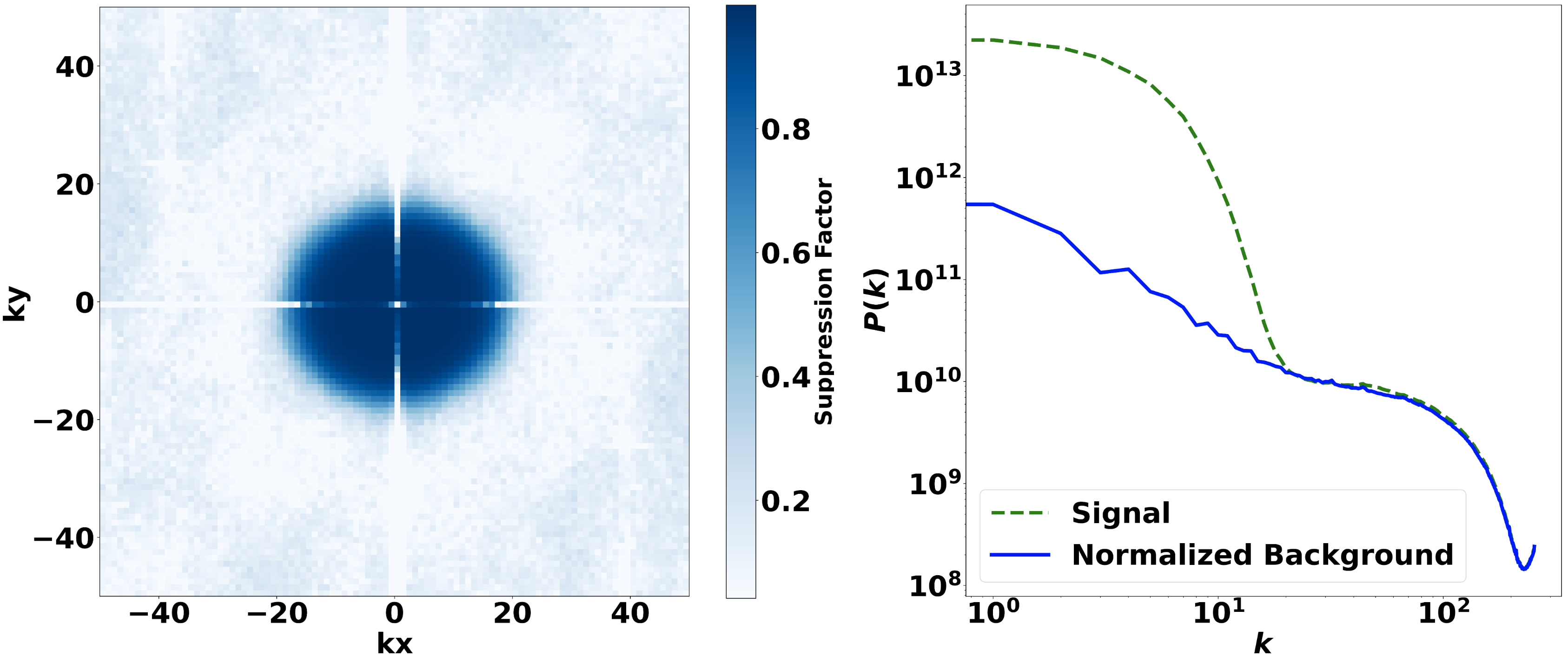}
\caption{Illustration of 2D Wiener filter and 1D Power Spectrum }
\label{fig:2DLowWiener}
\end{figure}

Using normalized background power spectrum and track power spectrum in Eq.~\ref{equation:Wiener1}, we construct the 2D Wiener filter shown in Fig.~\ref{fig:2DLowWiener} where~\textit{kx} and~\textit{ky} represent the suppression factor in the same coordinates as the 2D image power spectrum. 
The Wiener filter matrix contains values between 0 to 1;  any  negative value is due to random noise fluctuations in frequency space and is set to zero. 
It is illustrative to consider the 1D power spectrum,  obtained by radially averaging of 2D spectra of tracks and normalized background (Fig. \ref{fig:2DLowWiener}), right. 
It illustrates that the higher noise frequencies shape is common between the signal and background, whereas only signal events have large activity at lower frequencies. 
Therefore, it is primarily high frequencies that are suppressed by the filter.  
To apply the 2D filter to an image $I_{ij}$ and obtain a filtered image $Z_{ij}$, we apply Eq.~\ref{equation:Wiener1.4} 
\begin{equation}
    \centering
    Z_{ij}= FT^{-1}\left[FT(I)_{ij}W_{ij} \right].
    \label{equation:Wiener1.4}
\end{equation}
To further suppress noise at very high frequencies where the signal and background power spectra are similar, we also apply a loose additional low-pass filter, which has a negligible impact on the final images.  
An example showing the result of Wiener filtering on three Monte Carlo simulated beta events is shown in Fig.~\ref{fig:sim_betas}.

This filter is applied to all track events observed in the track imaging run, leading to clean and de-noised signal images.

\bibliographystyle{JHEP}
\bibliography{CRAB-0}

\end{document}